\RequirePackage{fix-cm}
\documentclass[final]{svjour3}                     
\smartqed  

\usepackage{graphicx}
\usepackage{amsmath}
\usepackage{amssymb}
\usepackage{setspace}
\usepackage[ruled,linesnumbered,vlined]{algorithm2e}
\usepackage{epsfig}
\usepackage{graphicx}
\usepackage{amsmath}
\usepackage{amssymb}
\usepackage{setspace}
\usepackage{color}
\usepackage{array}
\newcolumntype{C}[1]{>{\centering\let\newline\\\arraybackslash\hspace{0pt}}m{#1}}

\newcounter{exctr}

\begin{document}

\title{Fast Disk Conformal Parameterization of Simply-connected Open Surfaces}

\titlerunning{Disk conformal parameterization}        

\author{Pui Tung Choi \and Lok Ming Lui}



\institute{Pui Tung Choi \at
              Department of Mathematics, the Chinese University of Hong Kong \\
              \email{ptchoi@math.cuhk.edu.hk}           
            \and
            Lok Ming Lui \at
               Department of Mathematics, the Chinese University of Hong Kong\\
              \email{lmlui@math.cuhk.edu.hk}
}

\maketitle
\begin{abstract}

Surface parameterizations have been widely used in computer graphics and geometry processing. In particular, as simply-connected open surfaces are conformally equivalent to the unit disk, it is desirable to compute the disk conformal parameterizations of the surfaces. In this paper, we propose a novel algorithm for the conformal parameterization of a simply-connected open surface onto the unit disk, which significantly speeds up the computation, enhances the conformality and stability, and guarantees the bijectivity. The conformality distortions at the inner region and on the boundary are corrected by two steps, with the aid of an iterative scheme using quasi-conformal theories. Experimental results demonstrate the effectiveness of our proposed method.

\end{abstract}

\section{Introduction}
\label{intro}

Surface parameterization refers to the process of bijectively mapping a complicated surface to a simple canonical domain. In recent years, the use of parameterizations has been widespread in computer graphics and geometry processing. The applications of parameterizations include surface registration, texture mapping, mesh editing, remeshing, morphing, detail synthesis, mesh compression and medical visualization. For instance, in surface registration, which aims to find a one to one correspondence between two surfaces, it is common to parameterize the surfaces to simpler domains, such as the unit disk $\mathbb{D}$ or the unit sphere $\mathbb{S}^2$, to simplify the computation. This approach has been widely used in medical imaging for obtaining the surface registrations between anatomical structures, such as the cortical surfaces \cite{gu04,lui07,lam13} and the hippocampal surfaces \cite{lui10}. In texture mapping \cite{tutte02}, the geometric details and material properties are usually modeled as texture images. With the aid of the parameterizations of 3D meshes to the planar domain, the texture can be glued to the meshes. Besides, parameterizations are applied to solve PDEs on complicated 2D domains \cite{krichever04,krichever05,lui08}. The abovementioned applications reflect the importance of obtaining a good parameterization.

Numerous studies have been devoted to surface parameterizations. For simply-connected open surfaces, it is common to parameterize the surfaces to a unit disk. To achieve a meaningful parameterization, the distortion of certain geometric quantities, such as distance, area, and angle, should be minimized. By the Riemann mapping theorem, a simply-connected open surface is conformally equivalent to the unit disk $\mathbb{D}$. In other words, the existence of a disk conformal parameterization is theoretically guaranteed. Therefore, it is desirable to look for a disk conformal parameterization.

Conformal parameterization of disk-type surfaces has been a difficult topic in surface parameterization theory for a long time. Different research groups have developed brilliant algorithms to tackle the problem. Despite the effectiveness of the state-of-the-art approaches, there are still opportunities for further enhancements in the computational time and the conformality of the parameterizations. Firstly, as most of the latest algorithms are nonlinear, the computation is quite inefficient. This becomes an obstacle for practical applications in which a large number of surfaces are involved. Secondly, the conformality distortion is still far from negligible. The distortion affects the accuracy of the parameterizations, and thus hinders practical applications. There are two sources of the conformality distortion. One of the sources is the discretization of the surfaces and the operators in different algorithms. Since the surfaces are usually represented as triangulated meshes, the operators are discretized. Although the conformality of the parameterizations is theoretically guaranteed in the continuous case, certain numerical angular distortions inevitably exist for the discrete case under any algorithms. Another source is the limitations of the algorithms themselves due to different assumptions and conditions in the algorithms. In this work, we aim to develop a numerical method for disk conformal parameterizations that overcomes the above mentioned obstacles. First, we propose to speed up the computation by linearizing the algorithm as much as possible. To enhance the accuracy, we then propose a simple two-step iteration to correct the conformality distortion with the aid of quasi-conformal theories. Experimental results suggest that our proposed method outperforms other state-of-the-art approaches.

The rest of the paper is organized as follows. In Section \ref{previous}, we review the previous works in the literature related to our work. In Section \ref{contribution}, we outline the contributions of our work. Our proposed method is explained in details in Section \ref{main}. In Section \ref{implementation}, we describe the numerical implementation details of our proposed method. In Section \ref{experiment}, we show and analyze the experimental results of our proposed method. The paper is concluded in Section \ref{conclusion}.

\begin{table}[t]
    \centering
    \begin{tabular}{ |C{45mm}|c|c|c|c| }
    \hline
    Methods & Boundary & Bijectivity & Complexity\\ \hline
    Shape-preserving \cite{floater97}  & Fixed & Yes & Linear \\ \hline
    MIPS \cite{hormann00} & Free & Yes & Nonlinear \\ \hline
    ABF/ABF++ \cite{sheffer00,sheffer05} & Free & Local (no flips) & Nonlinear\\ \hline
    LSCM \cite{desbrun02,levy02} & Free & No & Linear\\ \hline
    Mean-value \cite{floater03} & Fixed & Yes & Linear\\ \hline
    Circle patterns \cite{kharevych05} & Free & Local (no flips) & Nonlinear \\ \hline
    Spectral conformal \cite{mullen08} & Free & No & Linear \\ \hline
    Double covering \cite{jin05} & Free & No & Nonlinear \\ \hline
    Discrete Ricci flow \cite{jin08}& Fixed & Yes & Nonlinear \\ \hline
    IDRF \cite{yang09} & Fixed & Yes & Nonlinear \\ \hline
    Yamabe Riemann map \cite{luo04} & Fixed & Yes & Nonlinear \\ \hline
    Holomorphic 1-form \cite{gu02} & Fixed & Yes & Nonlinear \\ \hline
    \end{tabular}
    \caption{Several previous works on conformal parameterization of disk-type surfaces.}
    \label{previouswork}
\end{table}

\section{Previous Works} \label{previous}

In this section, we describe some previous works closely related to our work.

Mesh parameterization has been extensively studied by different research groups. The goal is to map a complicated 3D or 2D surface to a simple parameter domain, such as the unit sphere $\mathbb{S}^2$ or the unit disk $\mathbb{D}$. Several surveys of mesh parameterization methods can be found in \cite{floater02,floater05,sheffer06,hormann07}.

Parameterizations inevitably create different kinds of geometric distortions. Thus, there is always a trade-off between the different types of distortions. This results in various criteria in determining the least distorted parameterizations. One of the criteria is the geodesic distance distortion. A parameterization that preserves distances is called \emph{isometric}. However, it is well known in differential geometry that distance preserving planar parameterizations only exist for developable surfaces, or equivalently, surfaces with zero Gaussian curvature \cite{docarmo76}. Hence, finding an isometric parameterization is impossible in most cases.

Another criteria is the area distortion. A parameterization without any area distortion is said to be \emph{authalic}. Desbrun et al. \cite{desbrun02} reported a discrete authalic parameterization method by introducing the Chi energy. Zou et. al. \cite{zou11} proposed an area preserving algorithm using Lie advection. Zhao et. al. \cite{zhao13} reported a method using optimal transport to achieve an area preserving mapping. However, as authalic parameterizations allow extreme angular distortions, they are less applicable in applications \cite{floater02}.

On the contrary, \emph{conformal} parameterization is more favorable as it preserves angles and hence the local geometry. For this reason, numerous studies have been devoted to conformal parameterizations. In particular, since conformal maps are equivalent to harmonic maps for genus-0 surfaces \cite{schoen97}, the problem can be turned into finding a harmonic map. The study of the discretization of harmonic maps is originated from \cite{pinkall93,eck95}. Pinkall and Polthier \cite{pinkall93} introduced a discretization of the Dirichlet energy for computing piecewise linear minimal surfaces. Eck et al. \cite{eck95} proposed a discrete harmonic mapping to approximate the continuous harmonic maps using finite element method.  Later, different conformal parameterization algorithms are proposed. Unlike the conformal parameterizations of genus-0 closed surfaces \cite{angenent99,haker00,gu02,gu04,hurdal04,lai13,lam13}, the presence of the surface boundary leads to more variations in the conformal parameterizations of simply-connected open surfaces. Floater \cite{floater97} introduced a method for making shape-preserving parameterizations of surface triangulations. Hormann and Greiner \cite{hormann00} presented the Most Isometric Parameterization of Surfaces method (MIPS) for disk-like surfaces. Sheffer and De Sturler \cite{sheffer00} proposed the Angle Based Flattening (ABF) method, which constructs a parameterization by minimizing a functional that punishes the angular distortion. In \cite{sheffer05}, Sheffer et. al. reported the ABF++ method, which is an extension of the ABF method that overcomes its drawbacks. In \cite{levy02}, L\'{e}vy et al. proposed a parameterization method by approximating the Cauchy-Riemann equations using the least-squares method. Desbrun et al. \cite{desbrun02} introduced the intrinsic parameterizations which minimize the distortion of different intrinsic measures of the surface patches. In \cite{floater03}, Floater derived a generalization of barycentric coordinates to improve methods for parameterizations. Kharevych et. al. \cite{kharevych05} introduced an approach for conformal parameterizations based on circle patterns. Gu and Yau \cite{gu02} constructed a basis of holomorphic 1-forms and integrated holomorphic differentials to obtain a conformal parameterization. Luo \cite{luo04} developed the combinatorial Yamabe flow for conformal parameterizations. Jin et. al. \cite{jin05} proposed a method for disk conformal parameterizations using the double covering \cite{gu03} followed by the spherical conformal mapping \cite{gu04}. Later, Mullen et. al. \cite{mullen08} reported a spectral approach to discrete conformal parameterizations. In \cite{jin08}, Jin et. al. proposed the discrete surface Ricci flow algorithm for conformal parameterizations. Yang et. al. \cite{yang09} generalized the discrete Ricci flow to improve the flexibility and robustness of the above method. The properties of the abovementioned conformal parameterization algorithms for disk-type surfaces are summarized in Table \ref{previouswork}.

As shown in Table \ref{previouswork}, for simply-connected open surfaces, there are two major types of conformal parameterizations, namely, 1) Free boundary parameterizations and 2) Fixed boundary parameterizations. Free boundary parameterizations do not restrict the shape of the boundary of the planar parameterizations. As there are more flexibilities on the boundary, less conformality distortions will be caused. However, because of the absence of boundary constraints, the planar parameterizations of two surfaces are usually completely different in shapes. This hinders the comparisons between different surfaces. For practical applications, it is desirable to obtain a planar parameterization with a more regular boundary. In this case, fixed boundary parameterization is preferred. The boundary of the mesh is usually restricted to a convex domain. Particularly, it is common to enforce the boundary to be a unit circle. Because of this extra constraint, the conformality distortion for fixed boundary parameterizations is unavoidably larger. This reflects the significance of finding a fast and accurate disk conformal parameterization algorithm.

\section{Contribution} \label{contribution}
The contributions of our work on disk conformal parameterizations are divided into three directions. Firstly, we improve the conformality distortion of the parameterizations by introducing a ``north pole-south pole'' iterative scheme. After obtaining an initial disk harmonic map, the conformality distortion of the inner region is corrected on the upper half plane by a composition of quasi-conformal maps. Then, we extend $\mathbb{D}$ to $\overline{\mathbb{C}}$ through a reflection, and correct the distortion near the boundary of $\mathbb{D}$ by a composition of quasi-conformal maps. Secondly, as every single step of our method is linear and the iteration converges shortly, our proposed algorithm is more computationally efficient than other state-of-the-art algorithms. Thirdly, our proposed method for disk conformal parameterizations is bijective. The bijectivity is ensured by the property of the Beltrami differential of the composition map. In summary, we propose an algorithm for disk conformal parameterization of genus-0 open surfaces with
\begin{enumerate}
 \item improved conformality;
 \item faster computation; and
 \item guaranteed bijectivity.
\end{enumerate}

\section{Proposed Method} \label{main}

In this section, we describe our proposed method for the disk conformal parameterizations of simply-connected open surfaces. The disk conformal parameterizations are achieved with the aid of an efficient iterative algorithm. In \cite{lam13}, the North Pole-South Pole iterative scheme was introduced for the fast spherical conformal parameterizations of genus-0 closed surfaces. The main idea of the iterative scheme is to improve the conformality distortions near the north pole and the south pole of the spherical parameterizations step by step. In the ``north pole'' step, a genus-0 closed surface is mapped to a unit sphere using a highly efficient method. The conformality distortion near the south pole of the sphere is small while the distortion near the north pole is relatively large. After that, in the ``south pole'' step, the conformality distortion near the north pole is corrected, with the region around the south pole kept fixed. In other words, to achieve a globally conformal parameterization, one can try to ensure the conformality of one part first, and then obtain the conformality of the other part in the second step, with the aid of the conformal part obtained before. Motivated by this idea, we introduce a ``North Pole-South Pole'' iterative scheme for disk conformal parameterizations of simply-connected open surfaces. In our case, instead of the actual geometric poles of the unit disk, by the ``north pole'' and the ``south pole'' we mean the two regions of a disk at which we handle the conformality distortion one by one. Table \ref{features} highlights the features of and the comparisons between the Fast Spherical Conformal Parameterization \cite{lam13} and our proposed method.

Our proposed method consists of three steps: 1) initialization, 2) ``north pole'' step, 3) ``south pole'' iteration. The three steps will be described in the subsequent three subsections respectively.

\begin{table}[t]
    \centering
    \begin{tabular}{ |C{22mm}|C{40mm}|C{40mm}| }
    \hline
    Features & Fast Spherical Conformal Parameterization \cite{lam13} & Our proposed method\\ \hline
    Type of surfaces & Genus-0 closed surfaces & Simply-connected open surfaces\\ \hline
    ``North pole'' step & Use the stereographic projection and work on $\mathbb{C}$ & Use the Cayley transform and work on the upper half plane\\ \hline
    ``South pole'' step & South pole stereographic projection & Reflection along the unit circle \\ \hline
    Boundary adjustment & No & Yes \\ \hline
    Output & Unit Sphere & Unit disk\\ \hline
    Bijectivity & Yes & Yes\\ \hline
    \end{tabular}
    \caption{Features of the Fast Spherical Conformal Parameterization \cite{lam13} and our proposed method.}
    \label{features}
\end{table}

\subsection{Initialization using the discrete harmonic map}

In the first step of our proposed method, we look for an initial map for the disk parameterizations. Among the existing algorithms, we use the disk harmonic map \cite{gu08} as an initialization since it is computationally efficient and easy to implement. We first briefly describe the harmonic map theory.

A map $f:M \to N$ between two Riemann surfaces is said to be \emph{conformal} if there exists a positive scalar function $\lambda$ such that $f^*ds_N^2 = \lambda ds_M^2$. It is easy to observe that every conformal map preserves angles.

The \emph{harmonic energy functional} for $f: M\to \mathbb{S}^2$ is defined as
\begin{equation}
E(f) = \int_M |\nabla f|^2 dv_M.
\end{equation}
In the space of mappings, the critical points of $E(f)$ are called \emph{harmonic mappings}. For genus-0 closed surfaces, conformal maps are equivalent to harmonic maps. For more details, please refer to \cite{schoen94,schoen97}. By Riemann mapping theorem, every simply-connected open surface $M$ can be conformally mapped onto $\mathbb{D}$. Also, a conformal map between two simply-connected open surfaces can be uniquely determined, provided that three-point correspondences are given.

In \cite{gu08}, Gu and Yau described a simple method to compute the disk harmonic map $f: M \to \mathbb{D}$ for a disk-type surface $M$. $f$ can be computed by solving the following Laplace equation:
\begin{equation}\label{Laplace}
\begin{cases}
 \Delta_{M} f(u) = 0 & \text{ if }u \in M\setminus \partial M\\
 f|_{\partial M} = g
\end{cases}
\end{equation}
where $g:\partial M \to \partial \mathbb{D}$ is given by the arc length parameterization.

In the discrete case, the Laplace equation $ \Delta_{M} f(u) = 0$ in Equation (\ref{Laplace}) becomes a sparse symmetric positive definite linear system. The boundary vertices $\{v_i\}_{i=0}^{n-1}$ are mapped to the unit circle according to the ratio of the edge lengths:
\begin{equation}
f(v_i) = (\cos \theta_i, \sin \theta_i),
\end{equation}
where $l_{[v_i,v_{i+1}]}$ denotes the length of the edge $[v_i,v_{i+1}]$ and
\begin{equation}
\left\{\begin{split}
s &:= \sum_{i=0}^{n-1} l_{[v_i,v_{i+1}]}\\
s_i&:= \sum_{j=0}^{i-1} l_{[v_j,v_{j+1}]}\\
\theta_i &:= 2 \pi \frac{s_i}{s}.
\end{split}\right.
\end{equation}
Hence, the Laplace Equation (\ref{Laplace}) can be efficiently solved. Note that the harmonic parameterization with fixed boundary condition is generally not conformal and induces conformality distortions. However, in practice, the harmonic disk parameterization is still a good enough mapping to serve as an initialization. The conformality distortions of the interior region and the boundary region of the disk will be corrected by the steps introduced in Subsection \ref{north} and \ref{south} respectively.

\subsection{Improvement of conformality on the upper half plane} \label{north}

\begin{figure*}[t]
    \centering
    \includegraphics[width=0.7\textwidth]{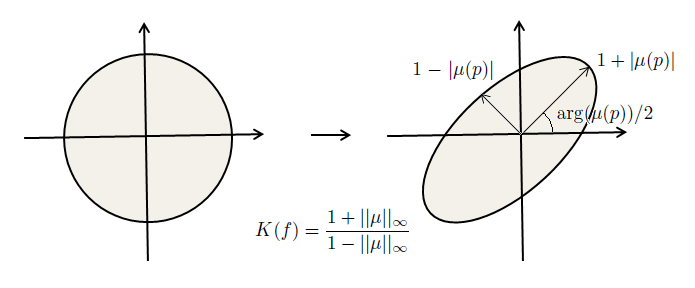}
    \caption{An illustration of how the conformality distortion can be determined by the Beltrami coefficient. The picture is adapted from \cite{fletcher06}.}
    \label{beltrami}
\end{figure*}

There are two drawbacks of the disk harmonic map algorithm \cite{gu08}. One of the drawbacks is that the conformality distortion is often quite large due to the restrictive circular boundary constraints. Secondly, the bijectivity is usually lost with bad triangulations. Foldings or overlaps may exist if there are extremely irregular triangles. We aim to alleviate these two drawbacks in this subsection and the following subsection.

To improve the conformality distortion of the initial disk harmonic map, our strategy is to compose the map with a \emph{quasi-conformal map}. Quasi-conformal maps are the generalizations of conformal maps, which are orientation preserving homeomorphisms between Riemann surfaces with bounded conformality distortions, in the sense that their first order approximations take small circles to small ellipses of bounded eccentricity \cite{gardiner00}. Mathematically, $f:\mathbb{C} \to \mathbb{C}$ is a \emph{quasi-conformal map} if it satisfies the Beltrami equation
\begin{equation}\label{beltramieqt}
\frac{\partial f}{\partial \bar{z}} = \mu(z) \frac{\partial f}{\partial z}
\end{equation}
for some complex-valued functions $\mu$ with $\| \mu \|_\infty<1$. $\mu$ is called the \emph{Beltrami coefficient} of $f$. Beltrami coefficient measures the conformality distortion of a map. In particular, $f$ is conformal around a small neighborhood of $p$ if and only if $\mu(p) = 0$.

\begin{figure*}[t]
    \centering
    \includegraphics[width=1\textwidth]{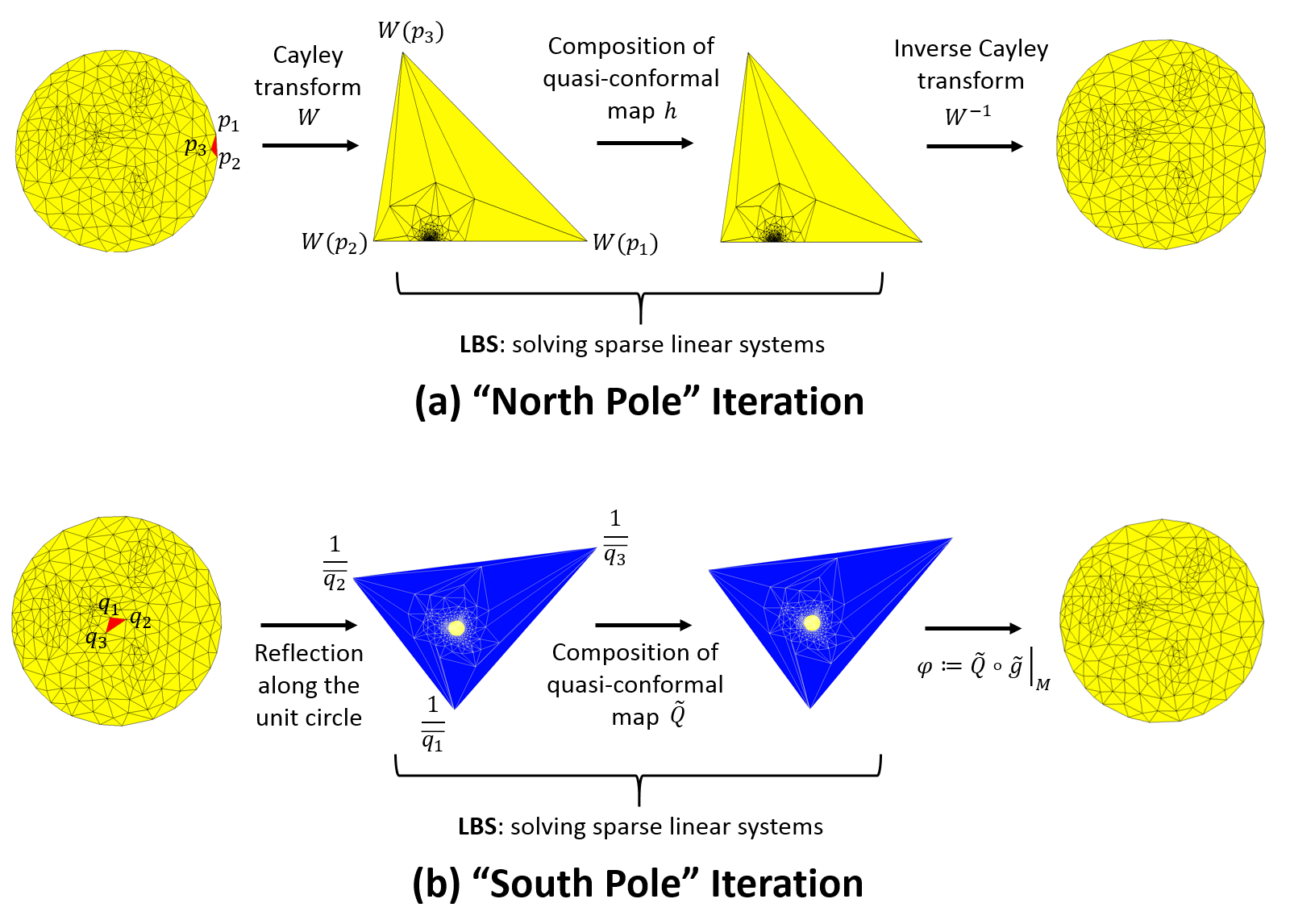}
    \caption{An illustration of the (a) ``north pole" iteration and (b) ``south pole" iteration.}
    \label{iterationillustration}
\end{figure*}

From $\mu(p)$, we can determine the angles of the directions of maximal magnification and shrinking and the amount of them as well. Specifically, the angle of maximal magnification is $arg(\mu(p))/2$ with magnifying factor $1 + |\mu(p)|$. The angle of maximal shrinking is the orthogonal angle $(arg(\mu(p)) - \pi)/2$ with shrinking factor $1 - |\mu(p)|$. Thus, the Beltrami coefficient $\mu$ gives us all the information about the properties of the map (See Figure \ref{beltrami}).

The maximal dilation of $f$ is given by:
\begin{equation}
K(f) = \frac{1+\|\mu\|_\infty}{1-\|\mu\|_\infty}.
\end{equation}

The Beltrami coefficient of a composition of quasi-conformal maps is related to the Beltrami coefficients of the original maps. Suppose $f: \Omega \to f(\Omega)$ and $g: f(\Omega) \to \mathbb{C}$ are two quasi-conformal maps with Beltrami coefficients $\mu_f$ and $\mu_g$ correspondingly. The Beltrami coefficient of the composition map $g \circ f$ is given by
\begin{equation}\label{composition_formula}
\mu_{g \circ f} = \frac{\mu_f+\frac{\overline{f_z}}{f_z} (\mu_g \circ f)}{1+\frac{\overline{f_z}}{f_z}  \overline{\mu_f} (\mu_g \circ f)}.
\end{equation}

Quasi-conformal map can also be defined between two Riemann surfaces. In this case, \emph{Beltrami differential} is used. A Beltrami differential $\mu(z)\frac{\overline{dz}}{dz}$ on a Riemann surface $S$ is an assignment to each chart $(U_\alpha, \phi_\alpha)$ of an $L_\infty$ complex-valued function $\mu_\alpha$, defined on local parameter $z_\alpha$ such that
\begin{equation}
\mu_\alpha \frac{d \overline{z_\alpha}}{dz_\alpha} = \mu_\beta \frac{d \overline{z_\beta}}{dz_\beta},
\end{equation}
on the domain which is also covered by another chart $(U_\beta, \phi_\beta )$. Here, $\frac{dz_\beta}{dz_\alpha} = \frac{d}{dz_\alpha}\phi_{\alpha\beta}$ and $\phi_{\alpha\beta} = \phi_\beta \circ \phi_\alpha$. An orientation preserving diffeomorphism $f: M \to N$ is called quasi-conformal associated with $\mu(z)\frac{\overline{dz}}{dz}$ if for any chart $(U_\alpha, \phi_\alpha)$ on $M$ and any chart $(U_\beta, \psi_\beta)$ on $N$, the mapping $f_{\alpha\beta} := \psi_\beta \circ f \circ f_\alpha^{-1}$ is quasi-conformal associated with $\mu_\alpha \frac{d \overline{z_\alpha}}{dz_\alpha}$. Readers are referred to \cite{gardiner00,fletcher06} for more details about quasi-conformal theories.

Beltrami differential is closely related to the bijectivity of a map. Specifically, if $f:M_1\to \mathbb{D}$ is a $C^1$ map satisfying $\|\mu_f\|_{\infty} <1$, then $f$ is bijective. This can be explained using the jacobian of $f$. The jacobian $J_f$ of $f$ is given by
\begin{equation}
J_f = \left| \frac{\partial f}{\partial z} \right|^2 (1-|\mu_f|^2).
\end{equation}
Since $\|\mu_f\|_{\infty} <1$, we have $\left| \frac{\partial f}{\partial z} \right|^2 \neq 0$ and $(1-|\mu_f|^2)>0$. Therefore, the Jacobian is positive everywhere. Since $\mathbb{D}$ is simply-connected and $f$ is proper, we can conclude that $f$ is a diffeomorphism. In fact, $f$ is a universal covering map of degree 1. Hence, $f$ must be bijective.

Intuitively, with a composition of two maps with the same Beltrami differential, the conformality distortion is cancelled out. In our case, let $\mu$ be the Beltrami differential of $f$, where $f: M \to \mathbb{D}$ is obtained by solving Equation (\ref{Laplace}). We proceed to look for a quasi-conformal map $g:\mathbb{D} \to \mathbb{D}$ with the same Beltrami differential $\mu$. The composition map, $\widetilde{f} := g\circ f^{-1}: M\to \mathbb{D}$, is then conformal. This can be explained by the following theorem:

\begin{theorem}\label{composition}
Let $f:M_1\to M_2$ and $g:M_2 \to M_3$ be quasi-conformal maps. Suppose the Beltrami differential of $f^{-1}$ and $g$ are the same. Then the Beltrami differential of $g\circ f$ is equal to 0. Hence, $g\circ f: M_1\to M_3$ is conformal.
\end{theorem}
\begin{proof}
Note that $\mu_{f^{-1}} \circ f = -(f_z / |f_z|)^2 \mu_f$. Since $\mu_{f^{-1}} =
\mu_g$, we have
\begin{equation}
 \mu_f+\frac{\overline{f_z}}{f_z} (\mu_g \circ f) = \mu_f+ \frac{\overline{f_z}}{f_z} (\mu_{f^{-1}} \circ f)
= \mu_f+\frac{\overline{f_z}}{f_z} (-\frac{f_z}{\overline{f_z}}) \mu_f = 0.
\end{equation}
Hence, by the composition formula,
\begin{equation}
\mu_{g \circ f} = \frac{\mu_f+\frac{\overline{f_z}}{f_z} (\mu_g \circ f)}{1+\frac{\overline{f_z}}{f_z} \overline{\mu_f}(\mu_g \circ f)} = 0.
\end{equation}
 Thus, $g \circ f$ is conformal.\qed
\end{proof}

Motivated by the above theorem, we can fix the conformality distortion of the initial $f$ by a quasi-conformal map $g:\mathbb{D}\to \mathbb{D}$. More specifically, suppose the Beltrami differential of $f$ is $\mu_f$, we compute another quasi-conformal map $g:\mathbb{D}\to \mathbb{D}$ with the same Beltrami differential. According to Theorem \ref{composition}, the composition map $g\circ{f}$ is conformal. The distortion of $f$ is therefore corrected.

Of course, one crucial issue is to efficiently compute $g$. Lui et al. \cite{lui13} proposed a linear algorithm, called the \emph{Linear Beltrami Solver({\bf LBS})}, to reconstruct a quasi-conformal map $g(x,y) = u(x,y) + i v(x,y)$ from its Beltrami coefficient $\mu_g = \rho + i\ \eta $ on rectangular domains in $\mathbb{C}$. Lam et al. \cite{lam13} extended this algorithm on triangular domains in $\mathbb{C}$. The brief idea of {\bf LBS} is as follows.

From the Beltrami Equation (\ref{beltramieqt}), we have
\begin{equation}
\mu_g = \frac{(u_x - v_y) + i(v_x + u_y)}{(u_x + v_y) + i(v_x - u_y)}.
\end{equation}
By direct computation, it remains to solve
\begin{equation}\label{eqt:BeltramiPDE}
\nabla \cdot \left(M \left(\begin{array}{c}
u_x\\
u_y \end{array}\right) \right) = 0\ \ \mathrm{and}\ \ \nabla \cdot \left(M \left(\begin{array}{c}
v_x\\
v_y \end{array}\right) \right) = 0
\end{equation}
where
\begin{equation}
M = \left( \begin{array}{cc} \frac{(\rho -1)^2 + \eta^2}{1-\rho^2 - \eta^2} & -\frac{2\eta}{1-\rho^2 - \eta^2}\\
-\frac{2\eta}{1-\rho^2 - \eta^2} & \frac{1+2\rho+\rho^2 +\eta^2}{1-\rho^2 - \eta^2} \end{array}\right).
\end{equation}
In the discrete case, solving the above elliptic PDEs (\ref{eqt:BeltramiPDE}), with certain boundary conditions on $u$ and $v$, can be discretized into solving a sparse symmetric positive definite linear system. Readers are referred to \cite{lui13,lam13} for details. For convenience, we denote the quasi-conformal map associated with the Beltrami differential $\mu$ by $\mathbf{LBS}(\mu)$.

In our case, we need to find a quasi-conformal map $g:\mathbb{D}\to \mathbb{D}$. This involves suitably allocating the boundary vertices on $\mathbb{S}^1$. In other words, instead of having a fixed boundary condition, we need to search for an optimal boundary correspondence $g|_{\partial \mathbb{D}}$ to reduce conformality distortions. This makes the problem nonlinear and causes computational difficulties. To alleviate this issue, our strategy is to transform the current domain to another domain, such that the problem can be linearized. More specifically, we use the Cayley transform to map the unit disk onto the upper half plane. The Cayley transform is a bijective conformal map. Mathematically, the Cayley transform $W: \mathbb{D} \to \mathbb{H}=\{x+iy | y\geq0; x,y \in \mathbb{R}\}$ is defined by
\begin{equation}
 W(z) = i\frac{1+z}{1-z}.
\end{equation}

Our problem is then transformed to finding a quasi-conformal map $h:\mathbb{H} \to \mathbb{H}$ whose Beltrami differential is equal to $\mu_{(W\circ f)^{-1}}$. According to Theorem \ref{composition}, the composition map $h\circ W \circ f: M\to \mathbb{H}$ is conformal.

Note that under the Cayley transform, the boundary of the disk is mapped onto the real axis $y=0$. In other words, to enforce a circular boundary, we only need to enforce that $h$ maps the real axis to the real axis. Equivalently, we only need to restrict $v=0$ on $\partial \mathbb{H}$ while solving equation (\ref{eqt:BeltramiPDE}) and put no restriction on $u$. This allows us to compute $h$ by solving two separate elliptic equations for $u$ and $v$.

In the discrete case, the surface $M$ is represented discretely by a triangulated mesh. The initial parameterization $f$ projects $M$ onto a triangulated mesh $\Omega$ of $\mathbb{D}$. $W$ maps $\Omega$ to a big triangle in $\mathbb{H}$. The three vertices of the big triangles are $W(p_1)$, $W(p_2)$ and $W(p_3)$, where $[p_1,p_2,p_3]$ is a triangular face of $\Omega$ enclosing the point $z=1\in \mathbb{D}$ (see Figure \ref{iterationillustration}(a)). To compute the desired quasi-conformal map $h$, we solve the Beltrami's equation, subject to the constraints that the three vertices of the big triangles are fixed and that vertices on the real axis slide along the real axis:

 \begin{equation}
  \begin{cases}
  h = \mathbf{LBS}(\mu_{(W \circ f)^{-1}})\\
  h(W(p_i)) = W(p_i)  \text{ for } i=1,2,3\\
  \textbf{Im} (h(W(z))) = 0 \text{ for any }z \in \partial \mathbb{D}.
  \end{cases}
 \end{equation}

The above can be formulated as two sparse symmetric positive linear systems, which can be solved efficiently using the conjugate gradient method.

After that, we map the upper half plane (or the big triangle in the discrete case) back to the unit disk using the inverse Cayley transform
\begin{equation}
W^{-1}(z) = \frac{z-i}{z+i}.
\end{equation}
As we enforce the boundary vertices to be on the real axis under the map $h$, the corresponding boundary vertices under the composition map
\begin{equation}\label{def_g}
 g:= W^{-1} \circ h \circ W \circ f
\end{equation}
will still be on $\partial \mathbb{D}$. This preserves the circular boundary condition of disk conformal parameterizations. Also, since $W^{-1}$ is conformal, the composition map $g$ is our desired disk conformal parameterization.

As a remark, the three vertices of the big triangle $\Omega$ correspond to the three vertices closest to $z=1$ on $\mathbb{D}$. Since the point $z=1$ here serves like the north pole in the stereographic projection in \cite{lam13}, we regard this step as the ``north pole'' step. It should be noted that the conformality distortion of the inner region of $\mathbb{D}$ is significantly improved by the composition of quasi-conformal maps, as explained in Theorem \ref{composition}. Also, the additional freedom on the boundary vertices slightly alleviates the conformality distortion near the boundary, although not perfectly. The conformality distortions near the boundary will be further adjusted in the next subsection. Besides, by the composition formula in Theorem \ref{composition}, we have $\|\mu_g\|_{\infty}<1$. Hence, $g$ is a diffeomorphism.

In summary, Figure \ref{iterationillustration}(a) gives a geometric illustration of the algorithm proposed in this subsection. To fix the conformality distortion at the inner region, our strategy is to compute a quasi-conformal map of the unit disk with the same Beltrami coefficient as the initial parameterization. In order to linearlize the computation of the quasi-conformal map, the unit disk is mapped to $\mathbb{H}$ or a big triangle in the discrete case by the Caley transform $W$. The problem is then reduced to solving the Beltrami's equation using $\mathbf{LBS}$ on the big triangle, which involves two sparse symmetric positive definite linear systems. The desired quasi-conformal map and hence the desired disk conformal parameterization can be obtained by an inverse Caley transform $W^{-1}$.

\subsection{Correction of boundary conformality distortion by reflection} \label{south}

After improving the conformality of the inner region, the next step is to correct the distortion near the boundary. Note that as we enforce the boundary of the parameter domain to be a unit circle (instead of a free boundary), this constraint causes conformality distortions near the boundary.

In \cite{lam13}, for spherical conformal parameterizations, the ``north pole'' iteration corrects the conformality distortion near the south pole of the sphere, and then the ``south pole'' iteration improves the conformality near the north pole, with the southernmost region fixed. Since the conformality distortion around the south pole is corrected by the ``north pole'' iteration, conformality distortions will not be induced by fixing the southernmost region in the ``south pole'' iteration. We extend this idea to our case. For our case, recall that the conformality of the inner region is significantly improved by the ``north pole'' step as described in subsection \ref{north}. Thus, the innermost region of the disk is the ``south pole'' we desire. Motivated by \cite{lam13}, we fix this least distorted region and correct the conformality distortion near the boundary by a quasi-conformal map.

Similar to the last subsection \ref{north}, we will make use of the $\mathbf{LBS}$ \cite{lui13} to construct the suitable quasi-conformal map. As observed in the last subsection, the composition of a quasi-conformal map can significantly reduce the distortion at the inner region but not near the boundary. Since our goal is to handle the distortion on the boundary, the aforementioned idea in the last subsection cannot be directly applied. To settle this problem, our strategy is to enlarge the domain, so that boundary region becomes the inner region of a much bigger domain. More precisely, we conformally reflect the unit disk along the circular boundary, such that the new domain of interest becomes the whole complex plane $\overline{\mathbb{C}}$. In the discrete case, the new domain is a big triangle whose three vertices are the reflected vertices of the triangle near the origin. The boundary region is now located at the inner region of the new big triangle (see Figure \ref{iterationillustration}b). Conformality distortion can be corrected by finding the appropriate quasi-conformal map of the big triangle.

Mathematically, Lui et. al. \cite{lui12} introduced an extension of a diffeomorphism on $\mathbb{D}$ to $\overline{\mathbb{C}}$ through a reflection as follows.

\begin{theorem} \label{extend}
 Let $f:\mathbb{D} \to \mathbb{D}$ be a diffeomorphism of the unit disk fixing 0 and 1 and satisfying the Beltrami equation $f_{\bar{z}} = \mu_f f_z$ with $\mu_f$ defined on $\mathbb{D}$. Then an extension of $f$ from $\mathbb{D}$ to $\overline{\mathbb{C}}$ given by
 \begin{equation}\label{f_extend}
  \tilde{f}(z) =
  \begin{cases}
  f(z) & \text{ if }|z| \leq 1 \\
  \frac{1}{\overline{f(1/\bar{z})}} & \text{ if } |z| > 1
  \end{cases}
 \end{equation}
satisfies the Beltrami Equation $\tilde{f}_{\bar{z}} = \tilde{\mu}_{\tilde{f}} \tilde{f}_z$ on $\overline{\mathbb{C}}$, where
 \begin{equation}\label{mu_extend}
  \tilde{\mu}_{\tilde{f}}(z) =
  \begin{cases}
  \mu_f(z) & \text{ if }|z| \leq 1 \\
  \frac{z^2}{\bar{z}^2}\overline{\mu_f(1/\bar{z})} & \text{ if } |z| > 1.
  \end{cases}
 \end{equation}
\end{theorem}

\begin{proof}
See \cite{lui12}.\qed
\end{proof}

In other words, by appropriately defining the Beltrami coefficient, one can extend a quasi-conformal map of $\mathbb{D}$ to a quasi-conformal map of $\overline{\mathbb{C}}$. 

Now we are ready to propose the ``south pole'' step for correcting the conformality distortion near the boundary of the disk parameterization. Using the formula of reflection in Equation (\ref{f_extend}), we construct a copy of the points on $\mathbb{D} \setminus \partial \mathbb{D}$ outside $\mathbb{D}$ by the correspondence
\begin{equation}
z \in \mathbb{D} \setminus \partial \mathbb{D} \longleftrightarrow \frac{1}{\bar{z}} \in \overline{\mathbb{C}} \setminus \mathbb{D},
\end{equation}
and extend the map $g$ in Equation (\ref{def_g}) to the extended map $\tilde{g}: \overline{\mathbb{C}} \to \overline{\mathbb{C}}$ defined by
\begin{equation}
\tilde{g}(z) = \begin{cases}
  g(z) & \text{ if }z \in \mathbb{D} \\
  \frac{1}{\overline{g(1/\bar{z})}} & \text{ if } z \in \overline{\mathbb{C}} \setminus \mathbb{D}.
  \end{cases}
\end{equation}
Recall that the conformality distortion of the innermost region of the disk parameterization is small after the ``north pole'' step. Since the outermost region of the new domain corresponds to the innermost region of $\mathbb{D}$ after the reflection (Equation (\ref{mu_extend})), the conformality distortion of the outermost region of the new domain is also negligible. With this characteristics, we apply the {\bf LBS} \cite{lui13} to compose the map $\tilde{g}$ by a quasi-conformal map $\widetilde{Q}:\overline{\mathbb{C}} \to \overline{\mathbb{C}}$ with the Beltrami differential $\tilde{\mu}_{{\widetilde{g}}^{-1}}$, leaving the outermost region of the new domain fixed:
 \begin{equation}
  \begin{cases}
  \widetilde{Q} = \mathbf{LBS}(\tilde{\mu}_{{\widetilde{g}}^{-1}})\\
  \widetilde{Q}(z) = z \text{ for } |z| \gg 1.
  \end{cases}
 \end{equation}

The conformality of $\tilde{g}$ will be significantly improved, according to the composition formula of Beltrami differentials in Equation (\ref{composition_formula}). In particular, the conformality distortion near the boundary of the original disk will be alleviated. Besides, by Theorem \ref{extend}, the region corresponding to $\mathbb{D}$ will be exactly mapped onto $\mathbb{D}$ under the map $\widetilde{Q}$. In other words, the boundary of the region is guaranteed to be a perfect circle in the continuous case. This results in a disk conformal parameterization
 \begin{equation}
  \varphi:= \widetilde{Q} \circ \widetilde{g}|_M ,
 \end{equation}
 with the conformality distortions at both the boundary region and the inner region corrected. Moreover, as the composition of quasi-conformal maps results in a zero Beltrami differential, we have $\|\mu_{\varphi}\|_{\infty}<1$. Hence, $\varphi$ is guaranteed to be bijective.

In the discrete case, the domain obtained using the reflection formula in Theorem \ref{extend} is a big triangle $\Omega$, where the outermost triangular faces of $\Omega$ corresponds to the innermost triangular face near the origin in $\mathbb{D}$ (see Figure \ref{iterationillustration}b). The computation of $\tilde{Q}$ is reduced to finding a quasi-conformal map of $\Omega$, which is solved efficiently using $\mathbf{LBS}$. Now, as the Beltrami differential is piecewise constant on each triangular face $T$ in $\mathbb{D}$, we cannot directly apply Equation (\ref{mu_extend}) to obtain the Beltrami differential $\tilde{\mu}_{{\widetilde{g}}^{-1}}(\widetilde{T})$ on the reflected triangular faces $\widetilde{T}$ on $\Omega\setminus \mathbb{D}$. Instead, we approximate $\tilde{\mu}_{{\widetilde{g}}^{-1}}(\widetilde{T})$ by
\begin{equation}
\tilde{\mu}_{{\widetilde{g}}^{-1}}(\widetilde{T}) = \frac{(\overline{z_1}^2/z_1^2 + \overline{z_2}^2/z_2^2 + \overline{z_3}^2/z_3^2)}{3} \overline{\mu_{g^{-1}}(T)},
\end{equation}
where $T = [z_1,z_2,z_3]$. This approximation unavoidably introduces numerical errors. Hence, the boundary of the inner region, which is the original disk, may not be transformed to a perfect circle under the composition. In this situation, we project the image boundary to the unit circle. That is, for any vertex $z\in \partial \mathbb{D}$,
\begin{equation}
z \mapsto \frac{z}{|z|}.
\end{equation}
Then we repeat the ``south pole'' step to extend the unit disk to the big triangle using the reflection formula in Equation (\ref{f_extend}) and perform the composition of quasi-conformal maps again until convergence.

\section{Numerical Implementation} \label{implementation}

In this section, we describe the numerical implementation of our proposed method for the disk conformal parameterization in details.

Firstly, we introduce the discretization of the disk harmonic map \cite{gu08} that we use as an initial map. Let $K$ be the triangulation of a genus-0 open surface $M$. Denote the edge spanned by two vertices $u,v$ by $[u,v]$. The discrete harmonic energy of $f:K \to \mathbb{D}$ is given by
\begin{equation}
E(f) = \sum_{[u,v] \in K} k_{uv} ||f(u)-f(v)||^2.
\end{equation}
Here $k_{uv} = \cot \alpha + \cot \beta$, where $\alpha,\beta$ are the angles opposite to the edge $[u,v]$. This is known as the cotangent formula.

From the above, the Laplace-Beltrami operator is discretized as
\begin{equation}
\Delta_M f(v_i) = \sum_{v_j \in N(v_i)} k_{v_i v_j} (f(v_j)-f(v_i))
\end{equation}
where $N(v_i)$ is the set of the vertices of the one-ring neighbors of the vertex $v_i$. Hence, the Laplace equation (\ref{Laplace}) becomes a sparse linear system in the form
\begin{equation}
Az = b
\end{equation}
where $A$ is a square matrix, $A(i,j) = k_{v_i v_j}$, $A(i,j) = -\sum_{v_j \in N(v_i)} k_{v_i v_j}$ subject to the arc-length parameterized boundary constraint. The linear system can be efficiently solved using the conjugate gradient method. Therefore, the initialization of our proposed method can be computed efficiently.

One important mathematical quantity in our proposed method is the Beltrami differential. In the discrete case, the computation of the Beltrami differentials between surfaces in $\mathbb{R}^3$ can be simplified to the computation of Beltrami coefficients on $\mathbb{C}$. This simplification is done as follows.

As the surfaces are represented as triangulated meshes, it is convenient to define the Beltrami coefficients on the triangular faces. Suppose $f=(u + i v) :K_1 \subset \mathbb{R}^2 \to K_2 \subset \mathbb{R}^2$ is an orientation preserving piecewise linear homeomorphism between two planar triangulated meshes. To compute the associated Beltrami coefficient $\mu_f$, which is a complex-valued function defined on each triangular face of $K_1$, we approximate the partial derivatives on every face $T_1$ on $K_1$.

Suppose $T_1$ on $K_1$ corresponds to another triangular face $T_2$ on $K_2$. The approximation of $\mu_f$ on $T_1$ can be achieved using the coordinates of the six vertices of $T_1$ and $T_2$. Specifically, suppose $T_1 = [a_1+i\ b_1, a_2+ i\ b_2, a_3 + i\ b_3]$ and $T_2 = [w_1, w_2, w_3]$, where $a_1,a_2,a_3,b_1,b_2,b_3 \in \mathbb{R}$, and $w_1,w_2,w_3 \in \mathbb{C}$. We approximate the Beltrami coefficient
\begin{equation}
\mu_f(z) = \frac{\partial f}{\partial \bar{z}} \left/ \frac{\partial f}{\partial z}\right.
\end{equation}
on $T_1$ by
\begin{equation}
\mu_f(T_1) = \frac{\frac{1}{2} \left(Dx + i\ Dy \right) \left( \begin{array}{c} w_1 \\ w_2 \\ w_3 \end{array} \right) }{\frac{1}{2} \left(Dx - i\ Dy \right) \left( \begin{array}{c} w_1 \\ w_2 \\ w_3 \end{array} \right) },
\end{equation}
where
\begin{equation}
Dx = \frac{1}{2 Area(T_1)} \left( \begin{array}{c} b_3-b_2 \\ b_1-b_3 \\ b_2-b_1 \end{array} \right)^t \ \ \mathrm{and}\ \
Dy = -\frac{1}{2 Area(T_1)} \left( \begin{array}{c} a_3-a_2 \\ a_1-a_3 \\ a_2-a_1 \end{array} \right)^t.
\end{equation}

More explicitly, the approximation of the Beltrami coefficient $\mu_f$ on $T_1$ is given by
\begin{equation}
\mu_f(T_1) = \frac{\left((b_3-b_2)w_1 + (b_1-b_3)w_2 + (b_2-b_1)w_3\right) - i\ \left((a_3-a_2)w_1 + (a_1-a_3)w_2 + (a_2-a_1)w_3\right)}{\left((b_3-b_2)w_1 + (b_1-b_3)w_2 + (b_2-b_1)w_3\right) + i\ \left((a_3-a_2)w_1 + (a_1-a_3)w_2 + (a_2-a_1)w_3\right)}.
\end{equation}

The above method can be extended to compute the Beltrami differential of a quasi-conformal map $g: K_1 \subset \mathbb{R}^2 \to K_2 \subset \mathbb{R}^3$. In this case, to compute the Beltrami differential between the corresponding triangular faces, we use a rigid motion $R$ to translate each triangular face of $K_2$ onto $\mathbb{R}^2$. Then we can use the abovementioned method to compute the Beltrami coefficient of the map $R \circ g$ on the triangular face $T_1$ on $K_1$. Since rigid motions are conformal, we have
\begin{equation}
\mu_R = 0.
\end{equation}
Hence, by the composition formula in Equation (\ref{composition_formula}),
\begin{equation}
\begin{split}
\mu_{R \circ g} &= \frac{\mu_g+\frac{\overline{g_z}}{g_z} (\mu_R \circ g)}{1+\frac{\overline{g_z}}{g_z}  \overline{\mu_g} (\mu_R \circ g)} \\
&= \frac{\mu_g+0}{1+0} \\
&= \mu_g.
\end{split}
\end{equation}
That is, the Beltrami differential of $g$ is equal to that of $R \circ g$. Therefore, the extension introduced above for the computation of Beltrami differentials is valid.

With the above extension, we can easily obtain the Beltrami differentials of the mappings in the ``north pole'' step and the ``south pole'' iteration in Subsection \ref{north} and \ref{south} respectively. Recall that Beltrami differentials are associated with quasi-conformal maps. Now, we look for an efficient method to compute the quasi-conformal map $f$ associated with a given Beltrami differential $\mu_f$. To achieve this, we apply the {\bf LBS} \cite{lui13} to reconstruct a quasi-conformal map from a given Beltrami differential with the three vertices of the big triangule fixed. We briefly explain the key idea for the discretization of the {\bf LBS} \cite{lui13}.

Note that the quasi-conformal map associated with a given Beltrami differential can be obtained by solving equation (\ref{eqt:BeltramiPDE}). {\bf LBS} aims to discretize equation (\ref{eqt:BeltramiPDE}) and reduce it to a linear system.

For each vertex $v_i$, let $N_i$ be the collection of neighborhood faces attached to $v_i$. Let $T = [v_i,v_j, v_k]$ be a face and $w_l = f(v_l)$ where $l=i,j$ or $k$. Suppose $v_l = g_l + i\ h_l$ and $w_l = s_l + i\ t_l$ ($l=i,j,k$). Assume further that the Beltrami differential of the face $T$ is denoted by $\mu_f(T) = \rho_T + i\ \eta_T$. It can be proved that equation (\ref{eqt:BeltramiPDE}) can be discretized into the following linear system:

\begin{equation}\label{eqt:linearB12}
\begin{split}
\sum_{T\in N_i}\frac{1}{Area(T)} \left\{ (h_j-h_k)[\alpha_1(T) a_T + \alpha_2(T) b_T]+(g_j-g_k)[\alpha_2(T) a_T + \alpha_3(T) b_T] \right\} = 0\\
\sum_{T\in N_i}\frac{1}{Area(T)} \left\{ (h_j-h_k)[\alpha_1(T) c_T + \alpha_2(T) d_T]+(g_j-g_k)[\alpha_2(T) c_T + \alpha_3(T) d_T] \right\} = 0
\end{split}
\end{equation}
where
\begin{equation}
\left( \begin{array}{cc} \alpha_1(T) & \alpha_2(T)\\
\alpha_2(T) & \alpha_3(T) \end{array}\right) = \left( \begin{array}{cc} \frac{(\rho_T -1)^2 + \eta_T^2}{1-\rho_T^2 - \eta_T^2} & -\frac{2\eta_T}{1-\rho_T^2 - \eta_T^2}\\
-\frac{2\eta_T}{1-\rho_T^2 - \eta_T^2} & \frac{1+2\rho_T+\rho_T^2 +\eta_T^2}{1-\rho_T^2 - \eta_T^2} \end{array}\right),
\end{equation}
and $a_T$, $b_T$, $c_T$ and $d_T$ are certain linear combinations of the $x$-coordinates and $y$-coordinates of the desired quasi-conformal map $f$. Hence, we can obtain the $x$-coordinate and $y$-coordinate function of $f$ by solving the linear system in equation (\ref{eqt:linearB12}). For more details, please refer to \cite{lui13}.

As the computations of the Beltrami differentials and the associated quasi-conformal maps (using {\bf LBS} \cite{lui13,lam13}) are both linear, the ``north pole'' step and the ``south pole'' iteration are highly efficient. Hence, our proposed method significantly speeds up the computation of the disk conformal parameterizations. The detailed implementation of our proposed method is described in Algorithm \ref{algorithm}.

\begin{algorithm}[h]
\KwIn{A simply-connected open mesh $M$, an energy threshold $\epsilon$.}
\KwOut{A bijective disk conformal parameterization $\varphi:M \to \mathbb{D}$.}
\BlankLine
Denote the boundary of $M$ as $\partial M = [v_0,v_1,...,v_n]$. Compute the edge lengths $l_{[v_i,v_{i+1}]}$ for $i=0,1,...,n$, where $v_{n+1} := v_0$\;
Obtain an initial disk parameterization $f:M \to \mathbb{D}$ by
$$
 \begin{cases}
  \sum_{v \in N(u)} k_{uv} (f(u)-f(v)) = 0  & \text{if } u\notin \partial M \\
  f(v_i) = (\cos \theta_i, \sin \theta_i)  & \text{if } u = v_i \in \partial M
  \end{cases}
$$where $s:= \sum_{i=0}^{n-1} l_{[v_i,v_{i+1}]}, s_i:= \sum_{j=0}^{i-1} l_{[v_j,v_{j+1}]}$ and $\theta_i := 2 \pi s_i/s$\;
Apply the Cayley transform $W:\mathbb{D} \to \mathbb{H}$ defined by $$W(z) = i\frac{1+z}{1-z}$$\;
Compute the Beltrami differential $\mu_{(W \circ f)^{-1}}$ of the map $(W \circ f)^{-1}$\;
Compute the quasi-conformal map $$h = \mathbf{LBS}(\mu_{(W \circ f)^{-1}})$$ with the boundary vertices $(W \circ f)(v_i)$ restricted on the real axis\;
Project the upper half plane to the unit disk by the inverse Cayley transform $$W^{-1}(z) = \frac{z-i}{z+i}.$$ Denote $\varphi:=W^{-1} \circ h \circ W \circ f$\;
\Repeat{$\text{mean}(|\mu_{\varphi^{-1}}|) - \text{mean}(|\nu|)<\epsilon$}{
Update $\nu$ by the Beltrami differential $\mu_{\varphi^{-1}}$ of the map $\varphi^{-1}$\;
By reflection, extend $\varphi^{-1}$ and $\mu_{\varphi^{-1}}$ on $\mathbb{D}$ to $\widetilde{\varphi}^{-1}$ and $\tilde{\mu}_{{\widetilde{\varphi}}^{-1}}$ on a big triangular domain $B$ using Equation (\ref{f_extend}) and Equation (\ref{mu_extend}). For each face $T=[z_1, z_2, z_3]$ on $\mathbb{D}$, define
$$\tilde{\mu}_{{\widetilde{\varphi}}^{-1}}(\widetilde{T}) = \frac{(\overline{z_1}^2/z_1^2 + \overline{z_2}^2/z_2^2 + \overline{z_3}^2/z_3^2)}{3} \overline{\mu_{\varphi^{-1}}(T)}$$\;
Compute the quasi-conformal map $$\widetilde{Q} = \mathbf{LBS}(\tilde{\mu}_{{\widetilde{\varphi}}^{-1}})$$ with the outermost vertices of $B$ fixed\;
Update $\varphi$ by the restriction $\widetilde{Q} \circ \widetilde{\varphi}|_M$\;
Project the boundary of $\varphi(M)$ onto the unit circle\;
}
\caption{Fast disk conformal parameterization}
\label{algorithm}
\end{algorithm}

\section{Experimental Results} \label{experiment}

In this section, we demonstrate the effectiveness of our proposed method using various 3D simply-connected open meshes. The meshes are freely available on the AIM@SHAPE Shape Repository \cite{aim@shape}. The algorithm is developed using MATLAB on Windows 7 platform. All experiments are performed on a PC with a 3.40 GHz CPU. In our experiments, the error threshold in Algorithm \ref{algorithm} is set to be $\epsilon=10^{-5}$.

\begin{figure}[h]
\begin{center}
   \includegraphics[width=0.35\textwidth]{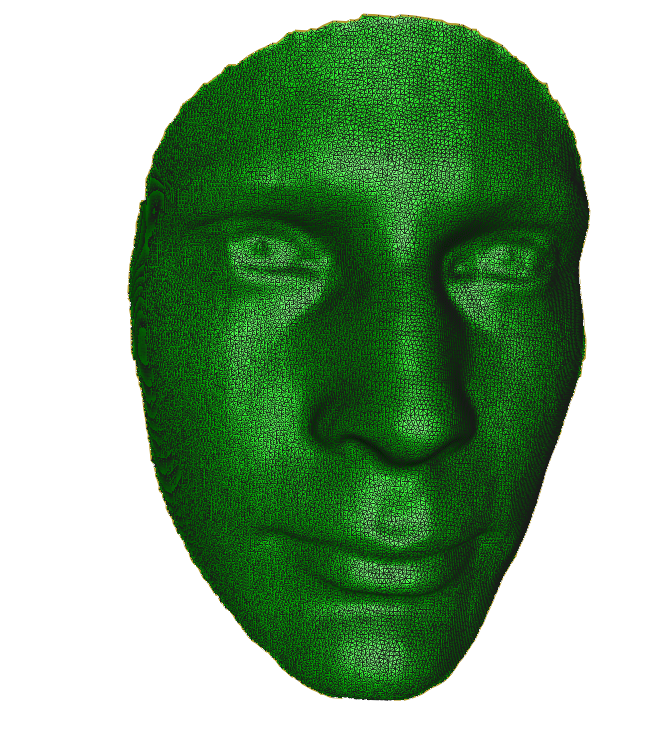}
   \includegraphics[width=0.35\textwidth]{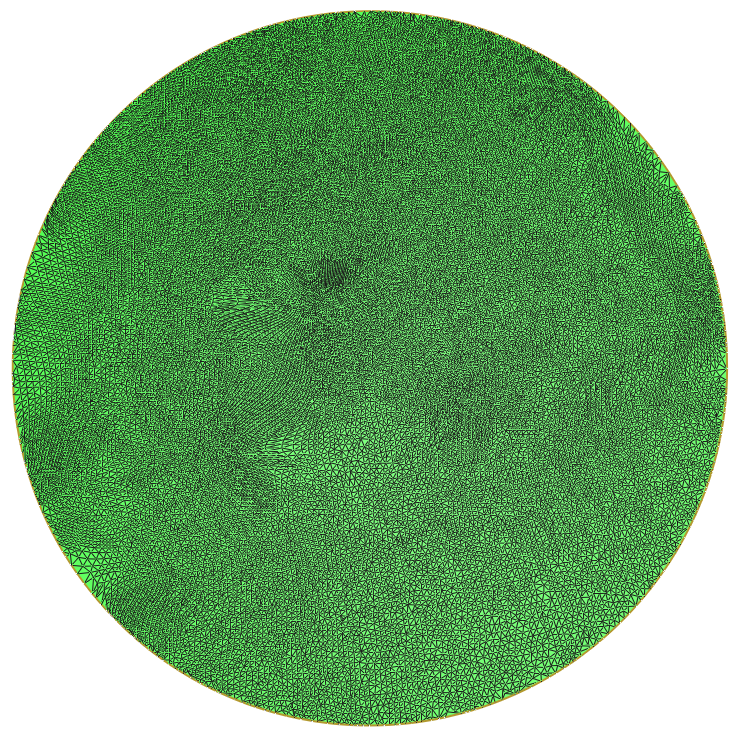}
   \includegraphics[width=0.35\textwidth]{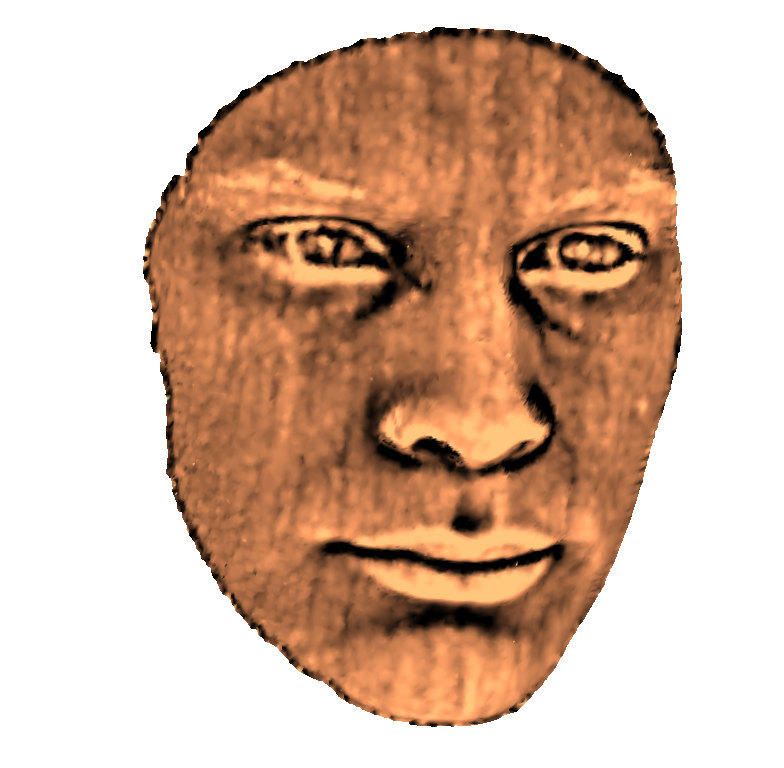}
   \includegraphics[width=0.35\textwidth]{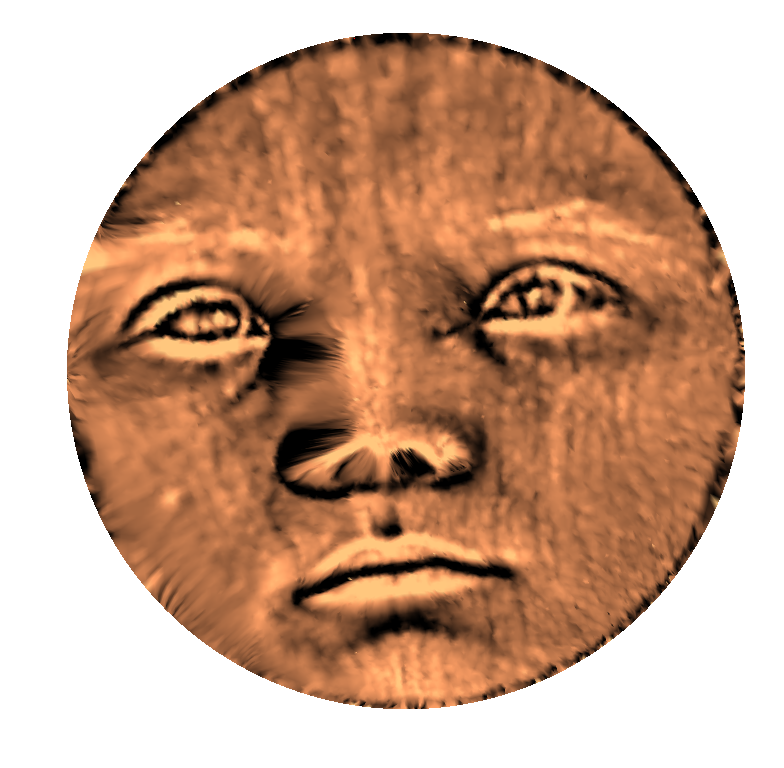}
 \end{center}
 \caption{A human face and its disk conformal parameterization using our proposed method. Top: the triangulations. Bottom: the mean curvature texture maps.}
 \label{fig:face}
 \end{figure}

\begin{figure}[h]
\begin{center}
   \includegraphics[width=0.35\textwidth]{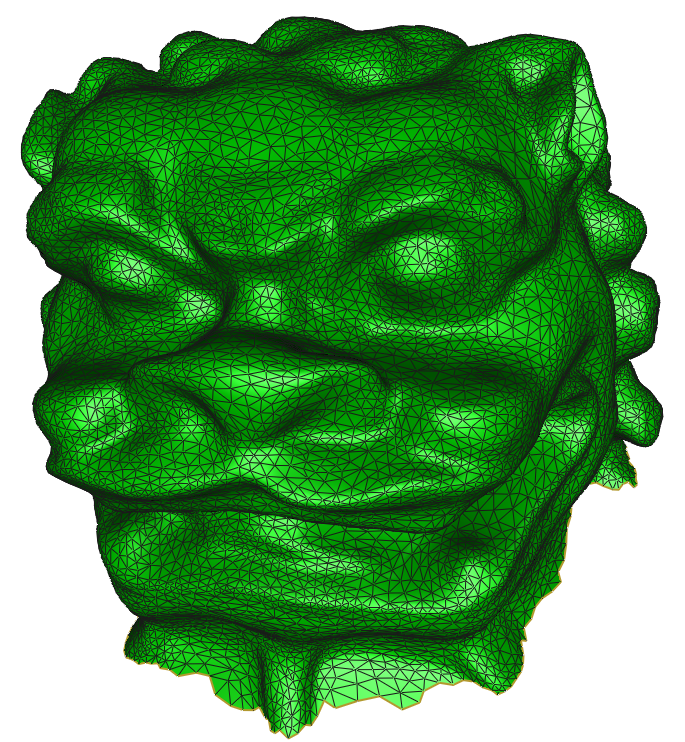}
   \includegraphics[width=0.35\textwidth]{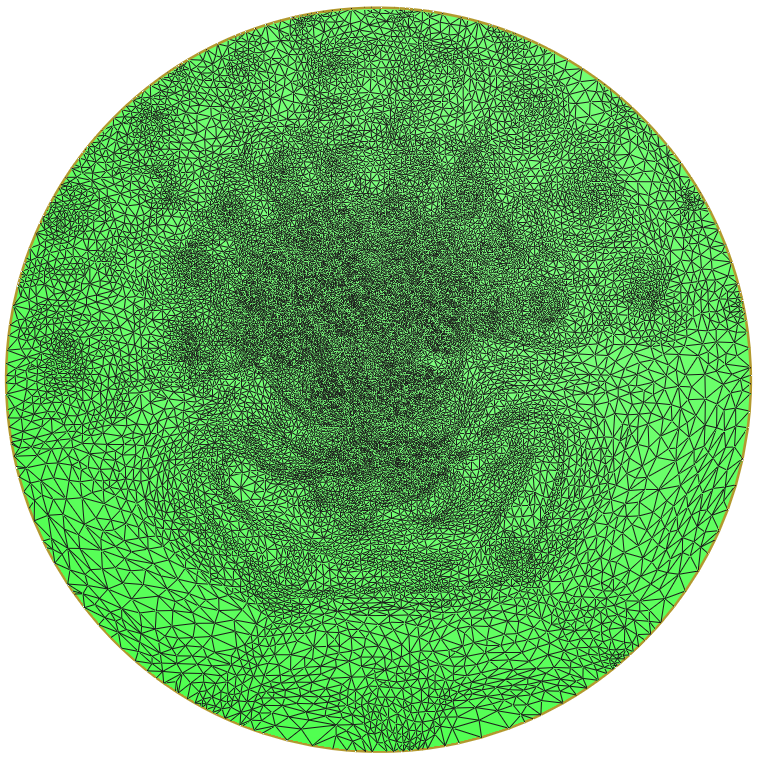}
   \includegraphics[width=0.35\textwidth]{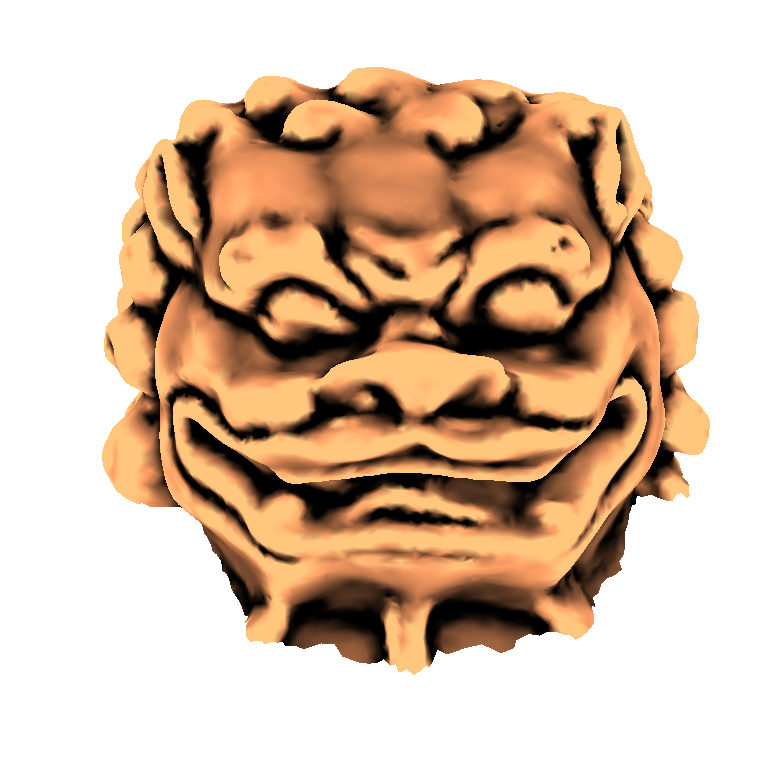}
   \includegraphics[width=0.35\textwidth]{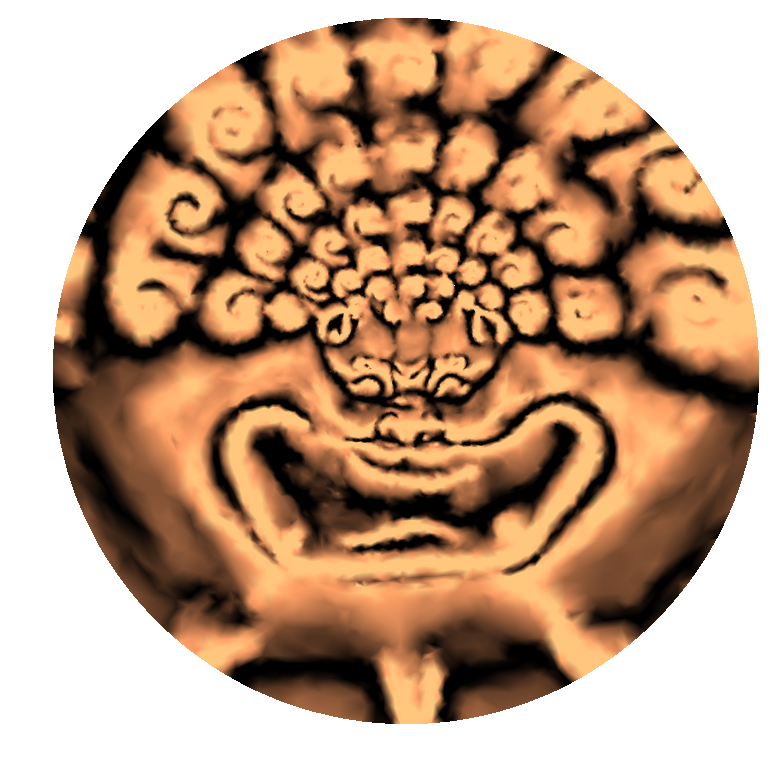}
  \end{center}
 \caption{A Chinese lion head and its disk conformal parameterization using our proposed method. Top: the triangulations. Bottom: the mean curvature texture maps.}
 \label{fig:lion}
 \end{figure}

\begin{figure}[h]
\begin{center}
   \includegraphics[width=0.35\textwidth]{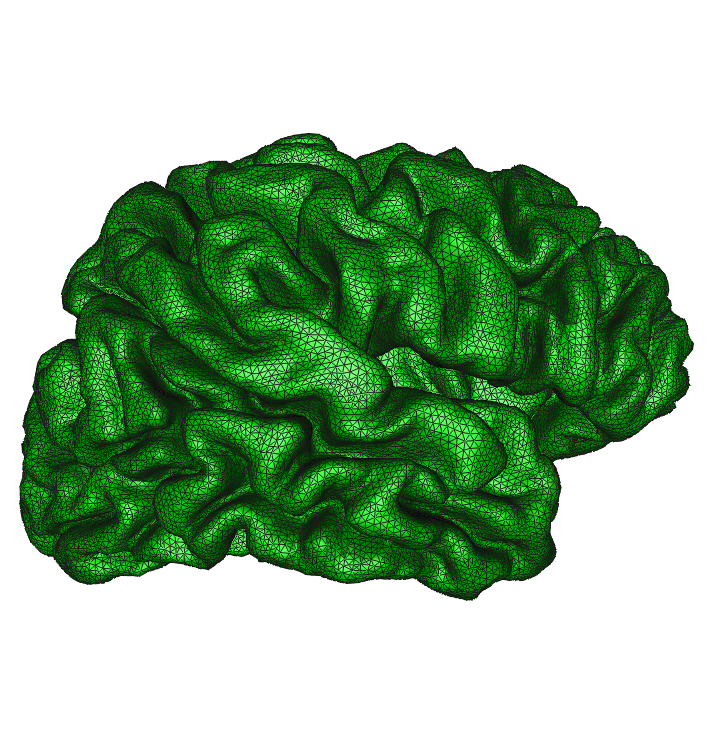}
   \includegraphics[width=0.35\textwidth]{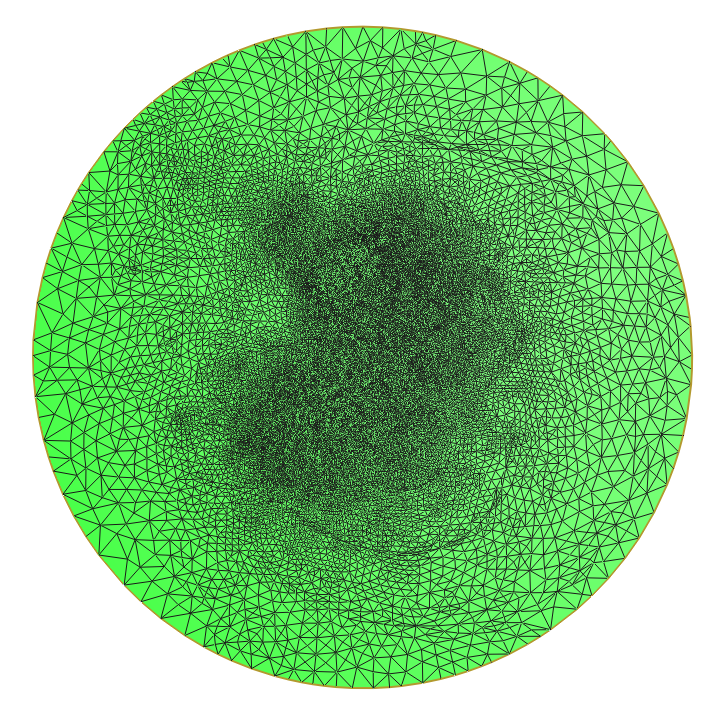}
   \includegraphics[width=0.35\textwidth]{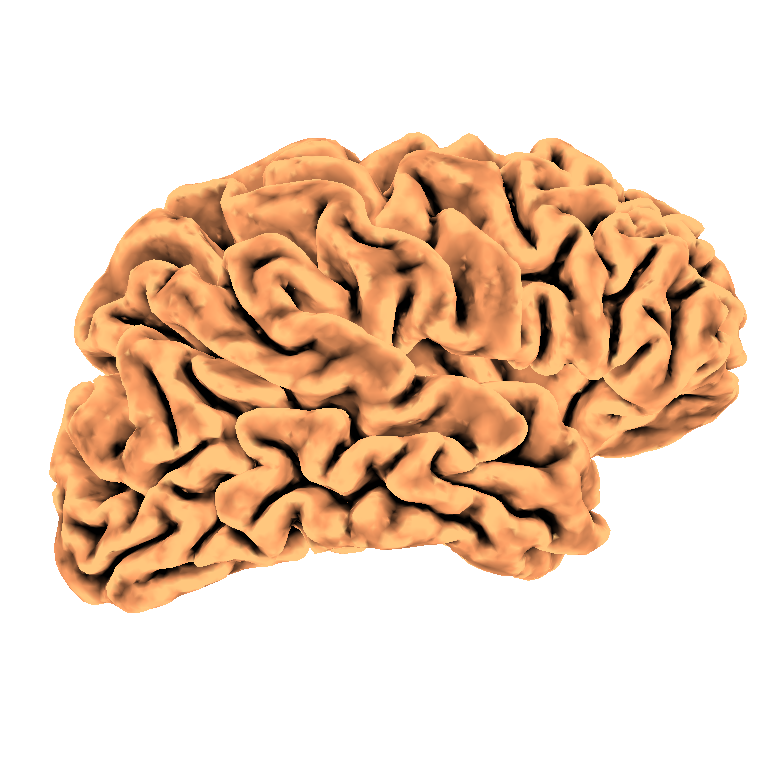}
   \includegraphics[width=0.35\textwidth]{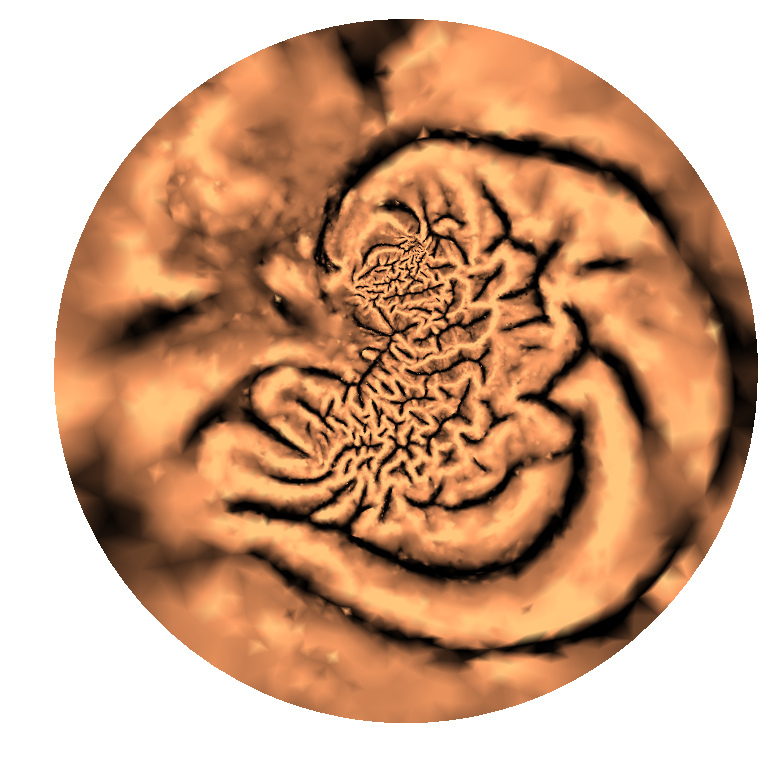}
  \end{center}
   \caption{A human brain and its disk conformal parameterization using our proposed method. Top: the triangulations. Bottom: the mean curvature texture maps.}
 \label{fig:brain}
 \end{figure}

\begin{figure}[h]
\begin{center}
   \includegraphics[width=0.32\textwidth]{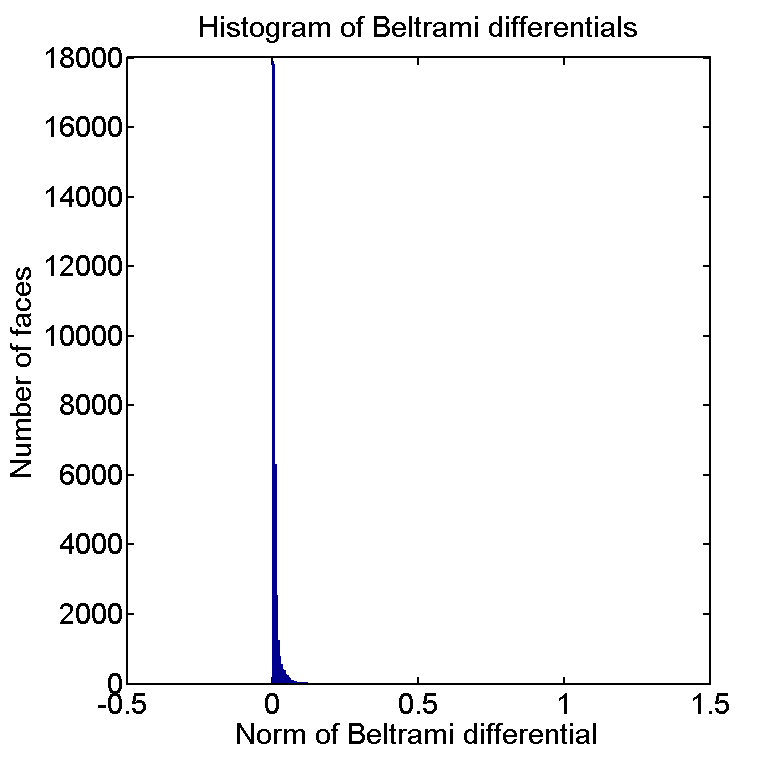}
   \includegraphics[width=0.32\textwidth]{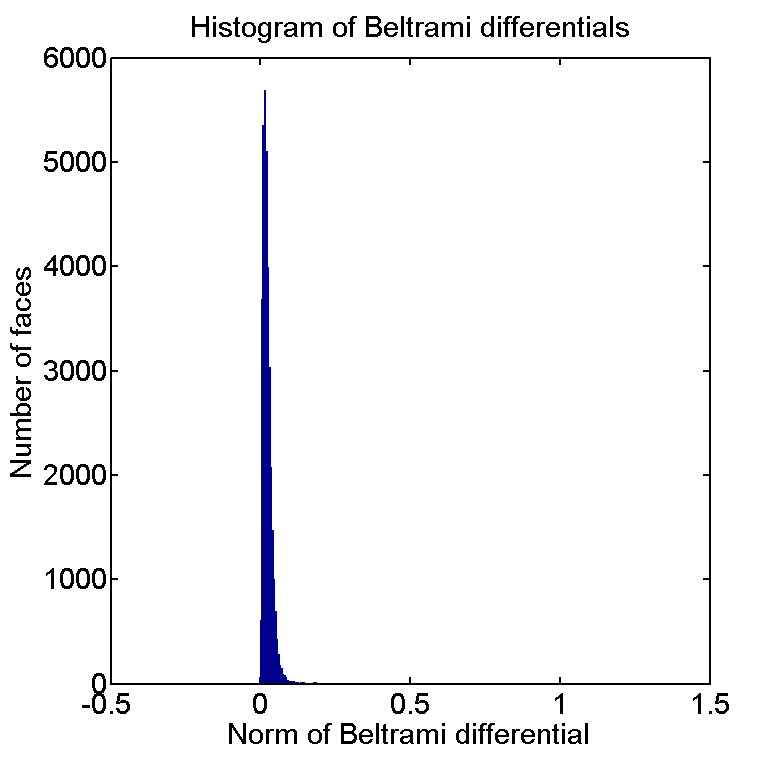}
   \includegraphics[width=0.32\textwidth]{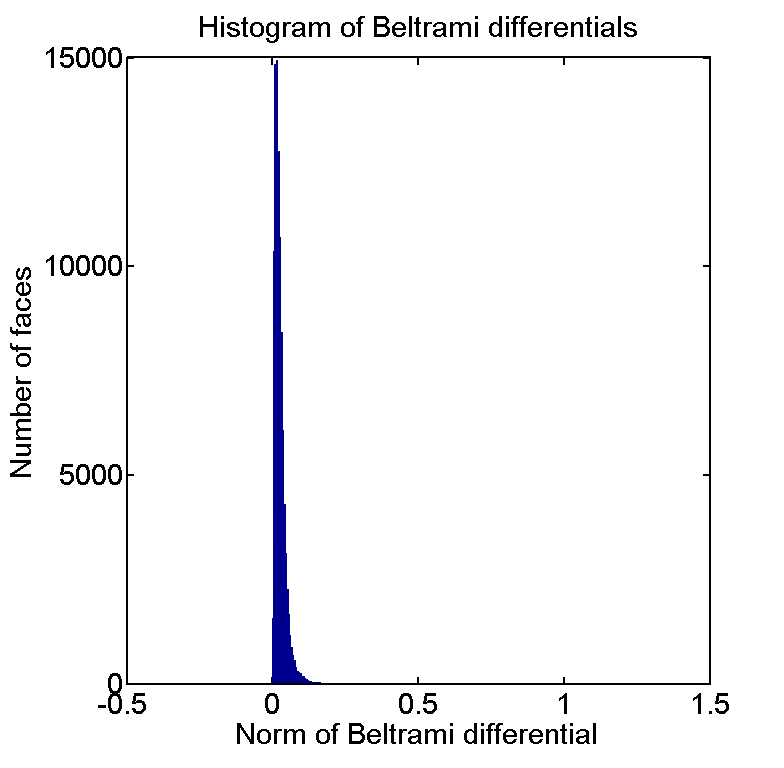}
 \end{center}
 \caption{Histograms of the norm of Beltrami differentials $|\mu|$ of our proposed method for a human face mesh, a Chinese lion head mesh and a human brain mesh. The small norms with the peak at $|\mu| \approx 0$ indicate that the conformality distortion is very small. Left: Human face. Middle: Chinese lion head. Right: Human brain.}
 \label{fig:histogram}
 \end{figure}

\begin{figure}[h]
\centering
   \includegraphics[width=0.32\textwidth]{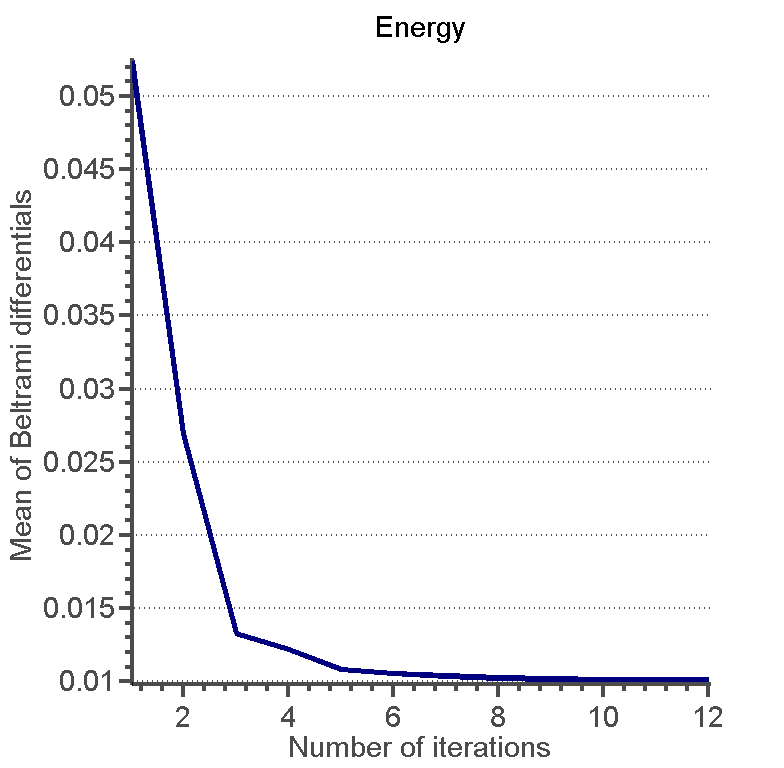}
   \includegraphics[width=0.32\textwidth]{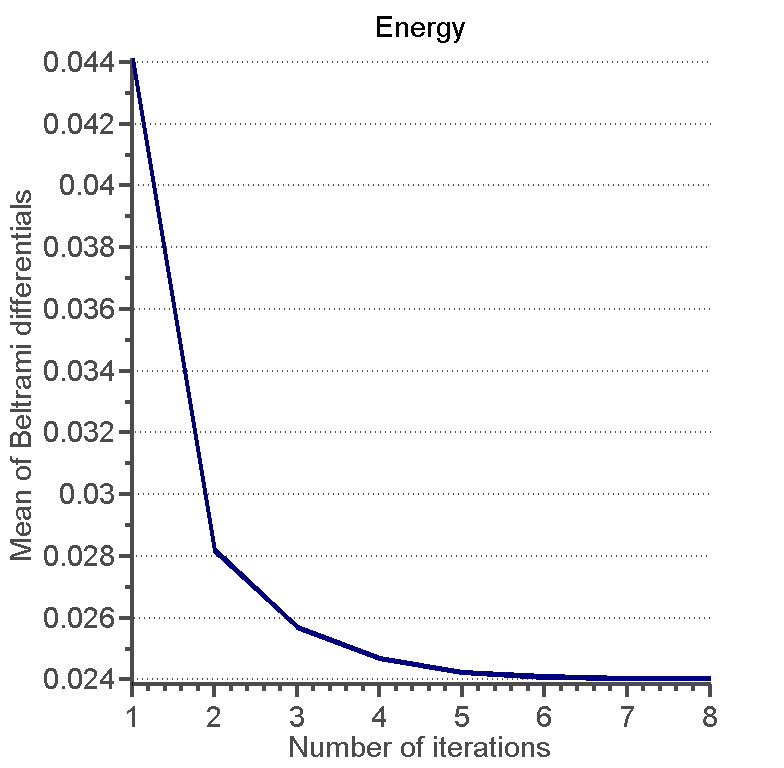}
   \includegraphics[width=0.32\textwidth]{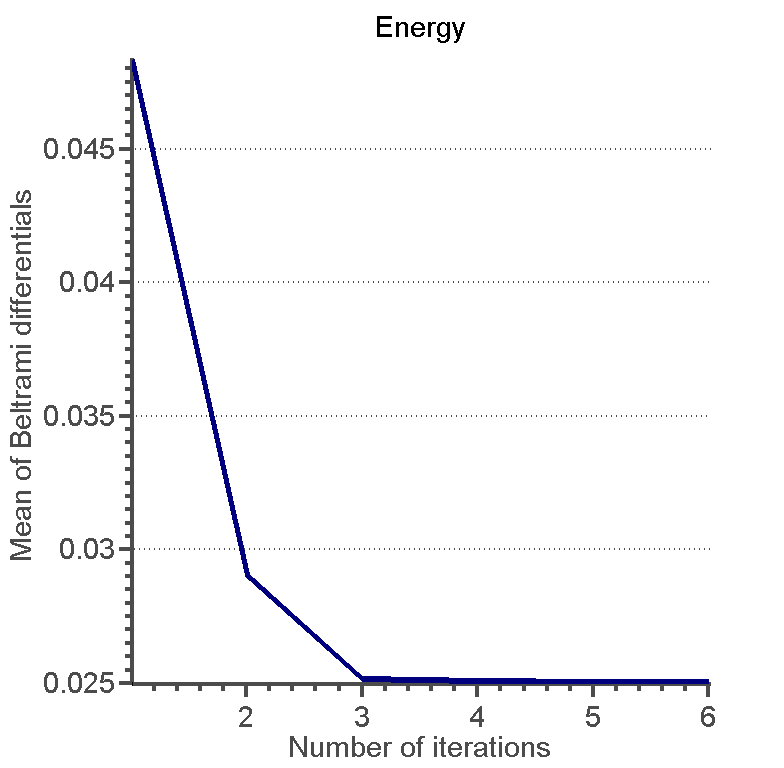}
 \caption{Energy plots of mean($|\mu|$) of our proposed method for a human face mesh, a Chinese lion head mesh and a human brain mesh. For all of the meshes, the iterations converge shortly. Left: Human face. Middle: Chinese lion head. Right: Human brain.}
 \label{fig:energy}
 \end{figure}

Figure \ref{fig:face}, Figure \ref{fig:lion} and Figure \ref{fig:brain} respectively show a human face mesh, a Chinese lion head mesh, a human brain mesh, and their disk conformal parameterizations obtained by our proposed method. The histograms of the norms of the Beltrami differentials are shown in Figure \ref{fig:histogram}. It is apparent that the peaks of the norms are close to $0$, which implies that the conformality distortions are small. Besides, from the energy plots shown in Figure \ref{fig:energy}, it can be observed that our proposed iterative method converges shortly. Hence, our method is very efficient.

To quantitatively assess the quality of our proposed method, three different factors are considered, including the computational time, the mean of the norm of the Beltrami differentials, and the standard deviation. The computational time evaluates the efficiency of our proposed method, the mean of the norm of the Beltrami differential checks if our proposed method is of small conformality distortions in general, and from the standard deviation we can see whether there exists any region with extremely large conformality distortions. As we aim at a bijective disk conformal parameterization, we compare our proposed method with four state-of-the-art algorithms that guarantee bijectivity and enforce a circular boundary. One of the methods is the discrete Ricci flow (RF) algorithm proposed by Jin et. al. \cite{jin08}. Another method is the inversive distance Ricci flow (IDRF) algorithm introduced by Yang et. al. \cite{yang09}. The Yamabe Riemann map algorithm by Luo \cite{luo04} and the holomorphic 1-form algorithm by Gu and Yau \cite{gu02} are also considered. The statistics of the performance of our proposed method and the four mentioned algorithms are listed in Table \ref{table_disk_para}.

\begin{table}
\centering
    \begin{center}
    \begin{tabular}{ |C{15mm}||c||c||c|c| }
    \hline
    Surfaces & No. of faces &  Our Method & RF \cite{jin08} & IDRF \cite{yang09}\\
    \cline{3-5}
    && \multicolumn{3}{ c| }{Time (seconds) / mean($|\mu|$) / sd($|\mu|$)}\\ \hline
    Human face & 49982 & 6.86/0.0101/0.0284 & 13.01/0.2421/0.1563 & fail \\ \hline
    Sophie & 41587 & 4.88/0.0056/0.0083 & 23.86/0.1476/0.0792 & 29.43/0.0057/0.0086 \\ \hline
    Max Planck & 99515 & 6.44/0.0102/0.0109 &  21.54/0.1431/0.0818 & 30.86/0.0103/0.0106 \\ \hline
    Mask & 62467 & 8.63/0.0043/0.0051 & 13.46/0.2379/0.1425 & fail\\ \hline
    Nicolo da Uzzano & 50042 & 5.71/0.0136/0.0314 & 10.34/0.2992/0.1434 & fail\\ \hline
    Julius Caesar & 433956 & 72.69/0.0032/0.0100 & 108.72/0.1033/0.0689 & 173.55/0.0033/0.0094\\ \hline
    Bimba & 48469 & 2.61/0.0217/0.0254 & 10.07/0.2947/0.1430 & fail\\ \hline
    Human brain & 96811 & 4.90/0.0250/0.0217 & 22.87/0.1861/0.1007 & 32.30/0.0249/0.0220 \\ \hline
    Hand & 105860 & 6.89/0.0194/0.0168 & 35.38/0.0550/0.0281 & 39.63/0.0211/0.0212\\ \hline
    Chinese lion & 34421 & 2.23/0.0240/0.0271 & 8.03/0.2029/0.1024 & 10.11/0.0238/0.0265 \\ \hline
    Lion vase & 98925 & 4.19/0.0236/0.0257 & 33.45/0.3687/0.1726 & fail\\ \hline
    \end{tabular}
    \end{center}

    \bigskip

    \begin{center}
    \begin{tabular}{ |C{15mm}||c||c||C{33mm}|C{33mm}| }
    \hline
    Surfaces & No. of faces &  Our Method & Yamabe Riemann map \cite{luo04} & Holomorphic 1-form \cite{gu02} \\
    \cline{3-5}
    && \multicolumn{3}{ c| }{Time (seconds) / mean($|\mu|$) / sd($|\mu|$)}\\ \hline
    Human face & 49982 & 6.86/0.0101/0.0284 & fail & 52.12/0.0111/0.0292 \\ \hline
    Sophie & 41587 & 4.88/0.0056/0.0083 & 34.32/0.0057/0.0085 & 57.36/0.0058/0.0083 \\ \hline
    Max Planck & 99515 & 6.44/0.0102/0.0109 & 34.34/0.0103/0.0106 & 92.74/0.0103/0.0109 \\ \hline
    Mask & 62467 & 8.63/0.0043/0.0051  & fail & fail\\ \hline
    Nicolo da Uzzano & 50042 & 5.71/0.0136/0.0314 & fail & 50.97/0.0143/0.0275\\ \hline
    Julius Caesar & 433956 & 72.69/0.0032/0.0100 & 175.01/0.0033/0.0094 & 462.63/0.0033/0.0096\\ \hline
    Bimba & 48469 & 2.61/0.0217/0.0254 & fail & 56.12/0.0230/0.0250\\ \hline
    Human brain & 96811 & 4.90/0.0250/0.0217 & 31.52/0.0249/0.0220 & 89.88/0.0251/0.0217\\ \hline
    Hand & 105860 & 6.89/0.0194/0.0168 & 44.25/0.0211/0.0212 & 104.04/0.0224/0.0269\\ \hline
    Chinese lion & 34421 & 2.23/0.0240/0.0271 & 14.82/0.0238/0.0265 & 40.43/0.0244/0.0271\\ \hline
    Lion vase & 98925 & 4.19/0.0236/0.0257 & fail & 137.93/0.0271/0.0282\\ \hline
    \end{tabular}
    \end{center}
    \caption{Performance of our proposed method and four state-of-the-art algorithms.}
    \label{table_disk_para}
\end{table}

\begin{figure}[h]
\begin{center}
   \includegraphics[width=0.35\textwidth]{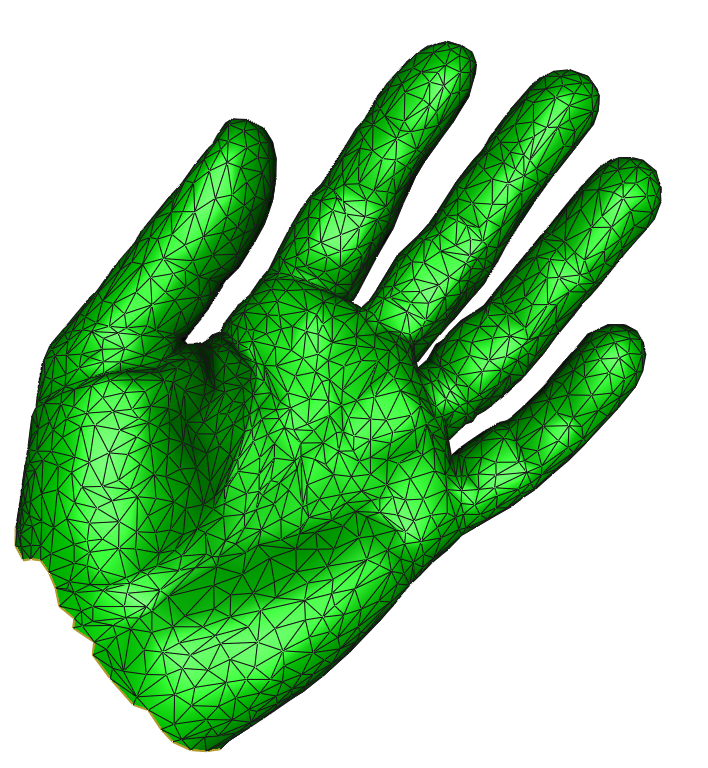}
   \includegraphics[width=0.35\textwidth]{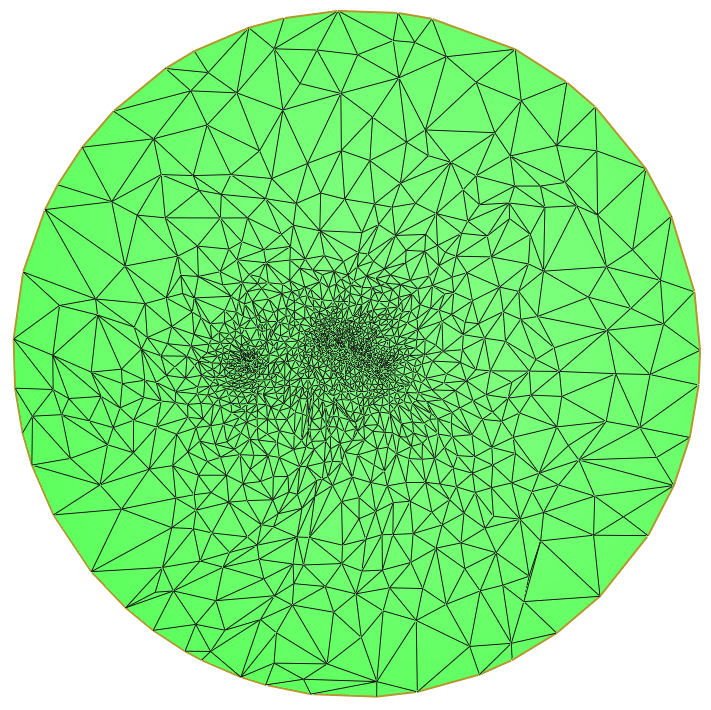}
   \includegraphics[width=0.35\textwidth]{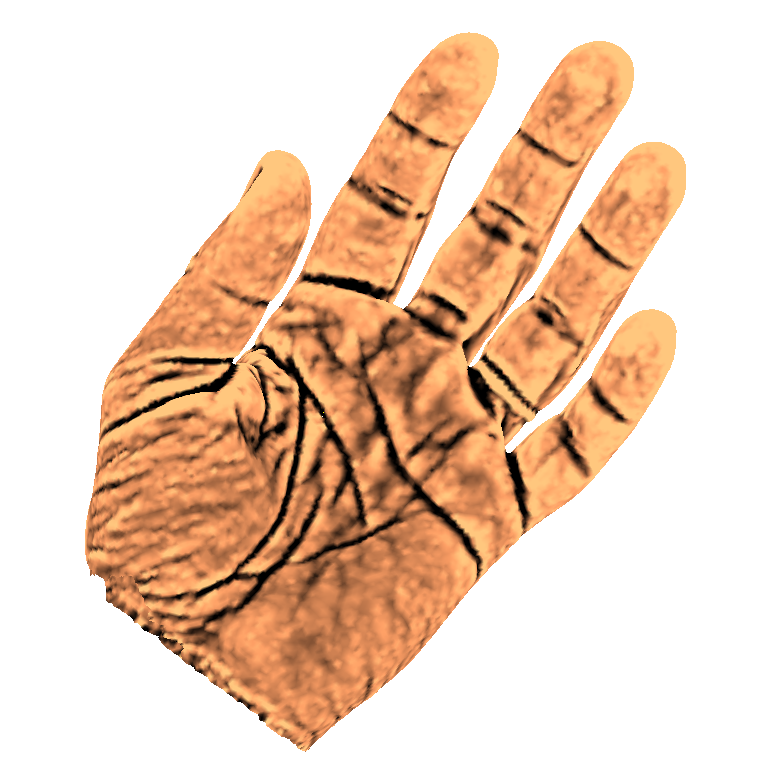}
   \includegraphics[width=0.35\textwidth]{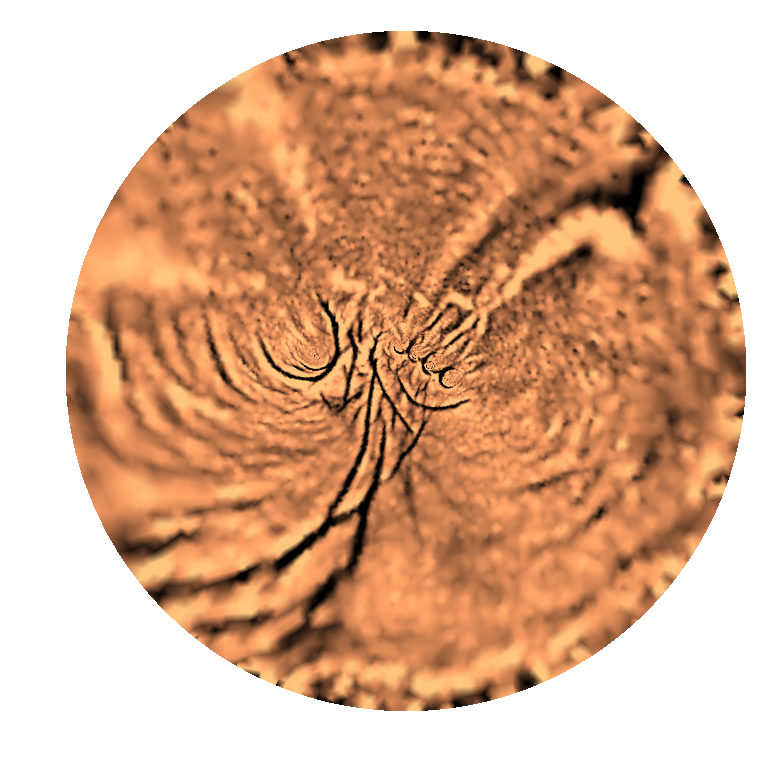}
\end{center}
 \caption{A hand mesh and its disk conformal parameterization using our proposed method. Top: the triangulations. Bottom: the mean curvature texture maps.}
 \label{fig:hand}
 \end{figure}

 \begin{figure}[h]
\begin{center}
   \includegraphics[width=0.4\textwidth]{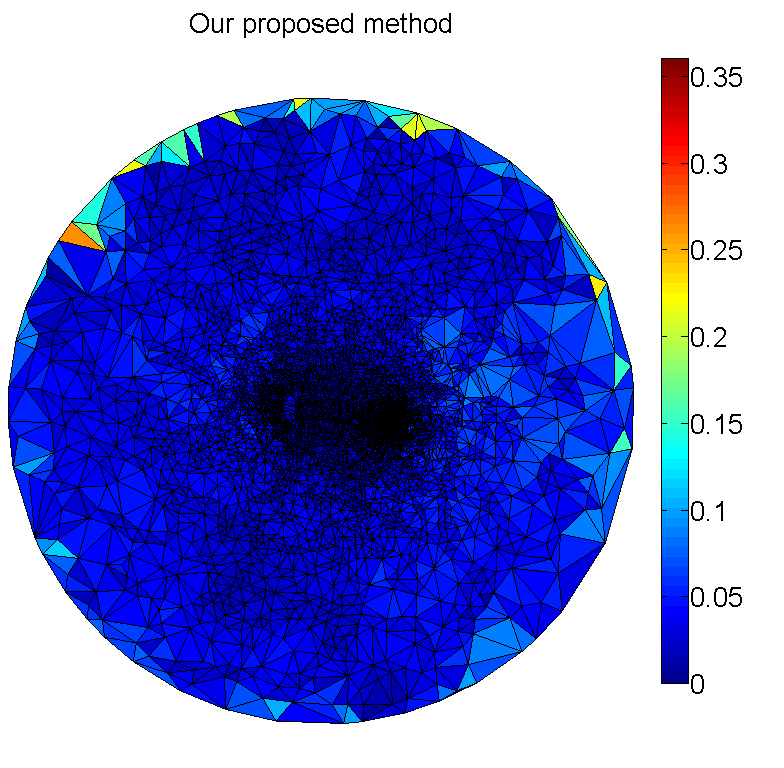}
   \includegraphics[width=0.4\textwidth]{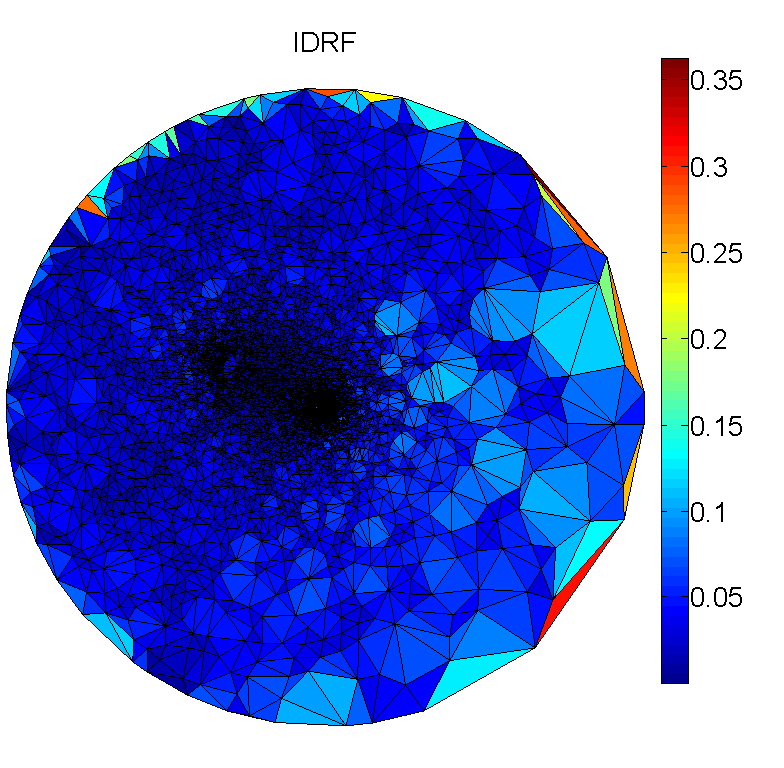}
 \end{center}
 \caption{Comparison of the norm of the Beltrami differentials between our proposed method and the IDRF algorithm \cite{yang09}. The colormaps show the norm of the Beltrami differentials on each triangular face. Left: Our proposed method. Right: the IDRF algorithm \cite{yang09}. }
 \label{fig:boundary}
 \end{figure}

\begin{figure}[h]
\begin{center}
   \includegraphics[width=0.4\textwidth]{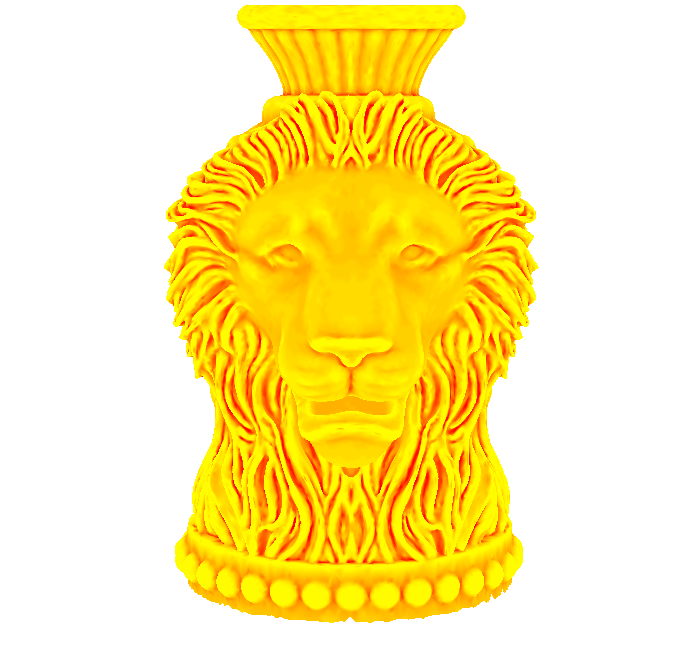}
   \includegraphics[width=0.4\textwidth]{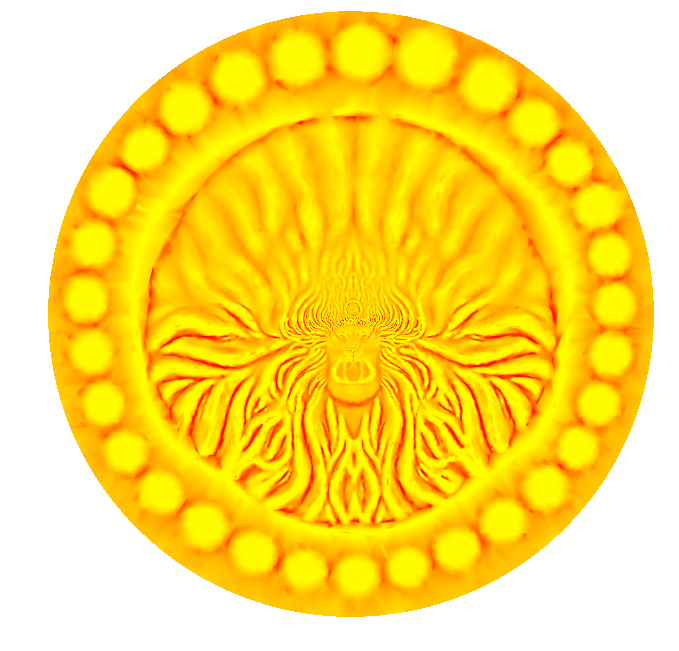}
 \end{center}
  \caption{A lion vase mesh and its disk conformal parameterization using our proposed method. The features of the lion vase mesh are well preserved by the disk parameterization. This demonstrates the conformality of the disk parameterization by our proposed method. }
 \label{fig:vaselion}
 \end{figure}

Our proposed method is highly efficient and accurate. For a 3D mesh with 100k faces, the time taken by our method is usually less than 10 seconds. Our method is capable of handling different types of meshes. Besides the typical human face meshes, our method also works for meshes with irregular shapes and bad triangulations. For instance, our proposed method successfully computes the disk conformal parameterization of a human hand mesh with long fingers (see Figure \ref{fig:hand}). As a remark, in all our experiments, the resulting disk conformal parameterizations are bijective. There exists no flips or overlaps in the disk conformal parameterizations obtained by our proposed method.

For a more detailed comparison, it is easy to see from Table \ref{table_disk_para} that the computational time of our method is shorter than that of the four aforementioned algorithms. Firstly, the computational time of our proposed method is 60\% shorter than that of the discrete Ricci Flow algorithm \cite{jin08} on average. For the conformality, it is apparent from the mean and the standard deviation of the norm of Beltrami differentials that our method surpasses the RF algorithm \cite{jin08}. The mean of the norm by our proposed method is over 85\% smaller than that by the RF algorithm \cite{jin08} on average. This shows that our proposed method has a better conformality. Also, the standard deviation by our proposed method is 80\% lower than that by the RF algorithm \cite{jin08} on average, which illustrates that the dispersion of the conformality distortion of our proposed method is much smaller.

Then, for the comparison with the Inversive Distance Ricci Flow algorithm \cite{yang09}, it is noteworthy that the computational time of our proposed method is 75\% shorter than that of the IDRF method \cite{yang09} on average. Besides, the conformality of our method is comparable to (and sometimes better than) that of the IDRF algorithm \cite{yang09}. In particular, the conformality near the boundary of our method is usually much better (see Figure \ref{fig:boundary}). This demonstrates the effectiveness of our proposed method in correcting the conformality distortion at the boundary region of the disk parameterization. More importantly, our method is applicable for a wider class of meshes. The IDRF algorithm \cite{yang09} may sometimes fail. For instance, the IDRF algorithm \cite{yang09} fails for the lion vase mesh shown in Figure \ref{fig:vaselion} while our proposed method works well. The features of the lion vase mesh, such as the circular patterns around the boundary and the texture of the hair, are well preserved. These results reflect the advantages of our proposed method.

\begin{figure}[h]
\begin{center}
   \includegraphics[width=0.4\textwidth]{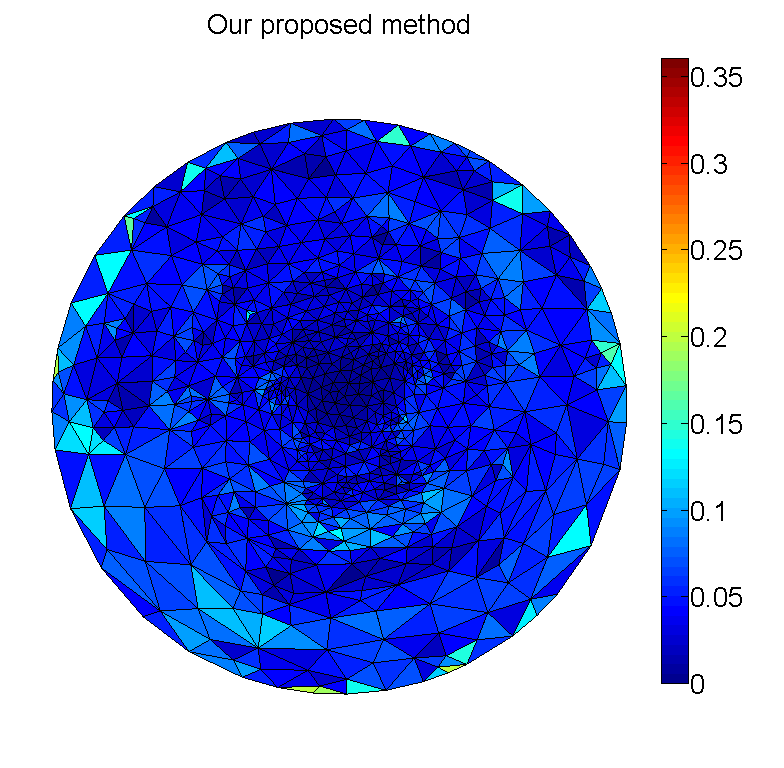}
   \includegraphics[width=0.4\textwidth]{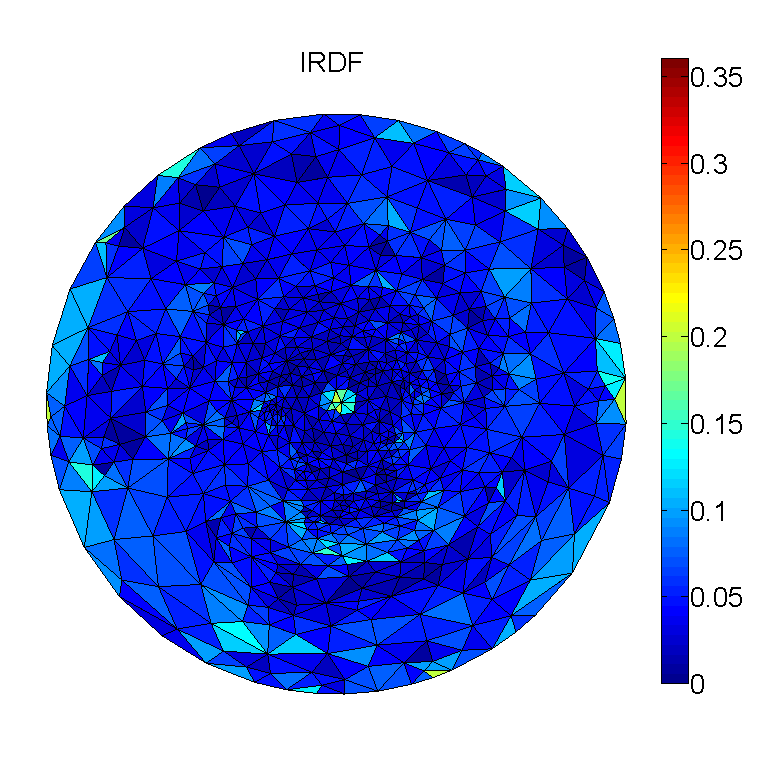}
   \includegraphics[width=0.4\textwidth]{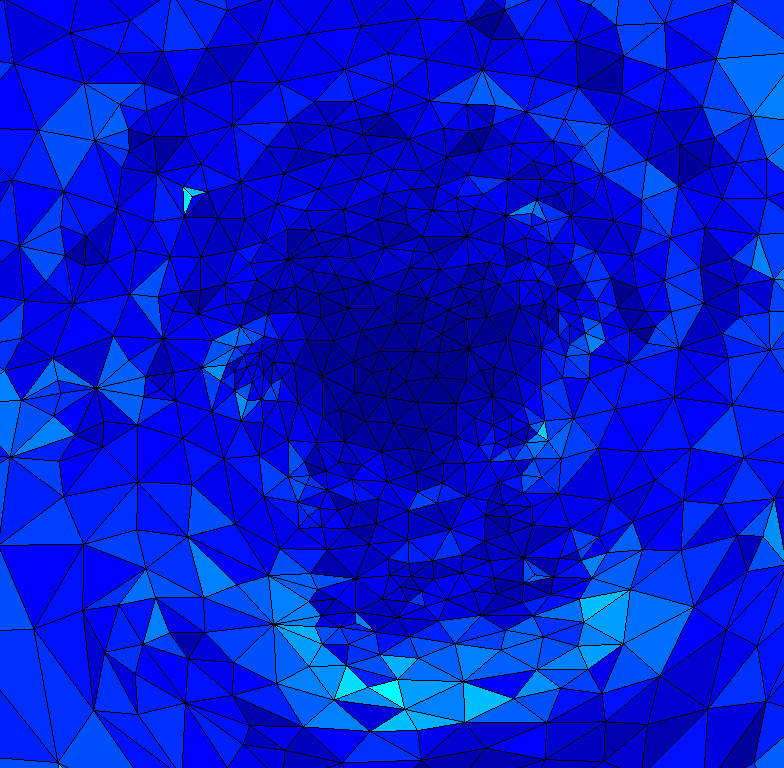}
   \includegraphics[width=0.4\textwidth]{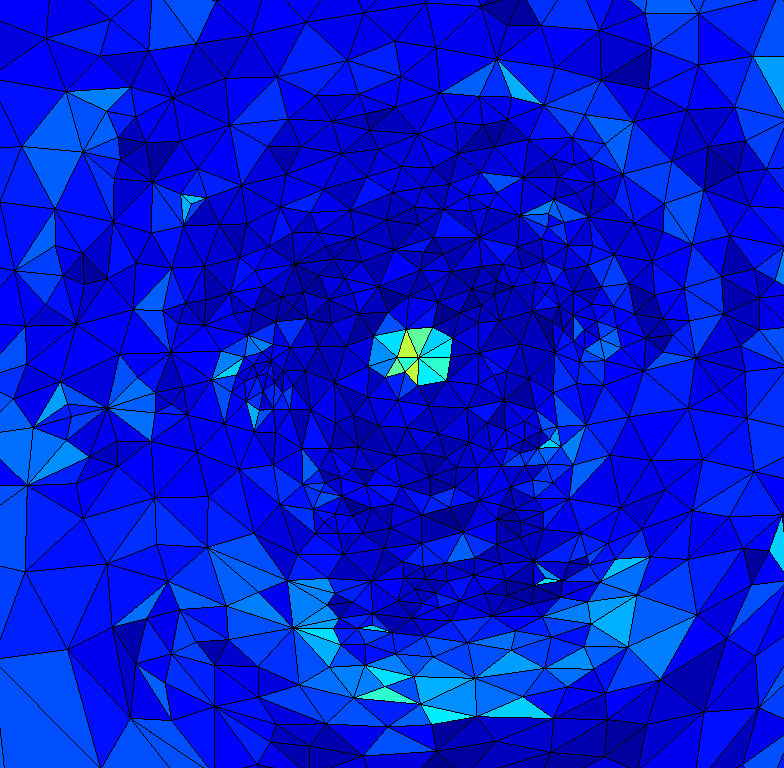}
\end{center}
 \caption{Comparison between the disk parameterizations obtained by our proposed method and by the IDRF algorithm \cite{yang09}. The colormaps show the norm of the Beltrami differentials on each triangular face. It is observed that there is a region with exceptionally large conformality distortion in the center of the disk by the IDRF algorithm \cite{yang09}. Top: the disk parameterizations obtained by our proposed method (left) and by the IDRF algorithm \cite{yang09}(right). Bottom: The zoom in of the center of the disks.}
 \label{fig:puncture}
 \end{figure}

When compared with the Yamabe Riemann map algorithm \cite{luo04}, our method also demonstrates a significant improvement in the computational time. More explicitly, our proposed method is 80\% faster than the Yamabe Riemann map algorithm \cite{luo04} on average. Our method is also applicable to a wider class of meshes as the Yamabe Riemann map algorithm \cite{luo04} fails for a number of meshes, especially for meshes with bad triangulations. For the remaining cases, the conformality of our method is comparable to (and sometimes better) than that of the Yamabe Riemann map algorithm \cite{luo04}.

For the differences between our proposed algorithm and the holomorphic 1-form algorithm \cite{gu02}, it is noteworthy that our proposed algorithm is 90\% faster than the holomorphic 1-form algorithm \cite{gu02} on average. The conformality of our method is comparable to (and often better) than that of the holomorphic 1-form algorithm \cite{gu02}. Besides, our method is more stable than the holomorphic 1-form algorithm \cite{gu02} as the holomorphic 1-form algorithm \cite{gu02} occasionally fails.

In addition, Figure \ref{fig:puncture} contrasts another feature between our proposed method and the IDRF algorithm \cite{yang09}. For the four abovementioned algorithms, one of the triangular faces has to be punctured at the beginning and filled at the end. For the region around the puncture, the conformality distortion is exceptionally large. On the contrary, the parameterization obtained by our proposed method is of small conformality distortion and is free of such unnaturally distorted regions. This again demonstrates the advantage of our proposed method.

Besides the four aforementioned methods, we also compare our proposed method with the double covering algorithm \cite{jin05} for simply-connected open surfaces. In \cite{jin05}, Jin et. al. suggested to glue two copies of the same surface along the boundaries to form a closed symmetric surface, and then computed the spherical conformal mapping \cite{gu04} of the new surface, and obtained the disk parameterization by applying the stereographic projection on the hemisphere. It is noteworthy that the spherical conformal mapping \cite{gu04} uses the Gauss map as initialization. In practice, it fails for most glued surfaces because the Gauss map is often undefined on the sharp ``boundaries'' of the glued surfaces. Instead of using the Gauss map, we use another method to obtain the initial spherical map. We apply the disk harmonic map \cite{gu08} and project the surface onto the southern hemisphere by the stereographic projection, and then glue the surface with a copy of it in the northern hemisphere to form the initial sphere. The energy threshold of the spherical conformal mapping \cite{gu04} is set to be $10^{-3}$. The comparison of the two algorithms is shown in Table \ref{table_double}.

\begin{table}
    \centering
    \begin{tabular}{ |C{15mm}|c|c|c|c| }
    \hline
    Surfaces & No. of faces & Our Method & Double covering \cite{jin05}\\
    \cline{3-4}
    && \multicolumn{2}{ c| }{Time (seconds) / mean($|\mu|$) / sd($|\mu|$) / $\sum_{\text{boundary}} |1-|z|^2|$}\\ \hline
    Human face & 49982 & 6.86/0.0101/0.0284/3.5305e-14 & 82.20/0.0679/0.0770/0.0036\\ \hline
    Sophie & 41587 & 4.88/0.0056/0.0083/4.3854e-14 & 31.62/0.0517/0.0390/0.0023\\ \hline
    Max Planck & 99515 & 6.44/0.0102/0.0109/1.2768e-14 & 115.53/0.0326/0.0214/0.0022\\ \hline
    Mask & 62467 & 8.63/0.0043/0.0051/1.0203e-13 & 74.88/0.1104/0.0461/0.0178\\ \hline
    Nicolo da Uzzano & 50042 & 5.71/0.0136/0.0314/3.5749e-14 & 99.71/0.0917/0.0624/0.0577\\ \hline
    Julius Caesar & 433956 & 72.69/0.0032/0.0100/1.3922e-13 & 467.42/0.1313/0.0361/0.0014\\ \hline
    Bimba & 48469 & 2.61/0.0217/0.0254/1.8763e-14 & 80.94/0.0638/0.0389/0.0625\\ \hline
    Human brain & 96811 & 4.90/0.0250/0.0217/9.3259e-15 & fail\\ \hline
    Hand & 105860 & 6.89/0.0194/0.0168/2.4314e-14 & fail\\ \hline
    Chinese lion & 34421 & 2.23/0.0240/0.0271/2.3204e-14 & 50.37/0.0477/0.0439/0.0034\\ \hline
    Lion vase & 98925 & 4.19/0.0236/0.0257/1.8430e-14 & 180.30/0.0959/0.0602/0.0101\\ \hline
    \end{tabular}
    \caption{Performance of our proposed method and the double covering method \cite{jin05}.}
   \label{table_double}
\end{table}

As illustrated by Table \ref{table_double}, the computational time of our proposed method is much shorter. Our proposed method is over 12 times faster than the double covering algorithm \cite{jin05} on average. Also, the conformality of our proposed method is significantly better. The mean and the standard deviation of the norms of the Beltrami differentials of our proposed method are 80\% and 60\% smaller than those of the double covering algorithm \cite{jin05} on average respectively. As a remark, the boundary of the disk parameterization obtained by the double covering algorithm \cite{jin05} is usually different from a perfect circle, while our proposed method always guarantees a circular boundary. This can be demonstrated from the last quantity in the table. The quantity measures the deviation of the boundary from a perfect circle, which is defined as $\sum_{\text{boundary}} |1-|z|^2|$. If the quantity is equal to zero, it means the boundary is a perfect circle. As shown in the table, our method successfully creates disk conformal parameterizations onto a perfect disk, while the double covering approach cannot.

\section{Conclusion and Future Work} \label{conclusion}
In this paper, we have presented a novel algorithm for the disk conformal parameterization of simply-connected open surfaces. Our method consists of three major steps. In the first step, we compute the initial disk parameterization using the disk harmonic map \cite{gu08}. Then, we project the unit disk to the upper half plane and compose the map with a quasi-conformal map to correct the conformality distortion at the inner region of the disk. After that, we extend the unit disk to a big triangular domain in $\mathbb{C}$ using a reflection, and compose the extended map with a quasi-conformal map with the extended Beltrami differential. Finally, by projecting the boundary to the unit disk if necessary and repeating the previous step, a disk conformal parameterization is obtained. It is noteworthy that the bijectivity of the parameterization is ensured by the composition formula of quasi-conformal maps. Experimental results have illustrated the effectiveness of our proposed method, with significant improvements in the computational time, the conformality and the stability. In the future, we will investigate the possibility of extending the proposed method for enhancing the parameterizations of high genus surfaces.




\end{document}